\documentclass{fundam}

\publyear{22}
\papernumber{2102}
\volume{185}
\issue{1}

\usepackage{url} 
\usepackage[ruled,lined]{algorithm2e}
\usepackage{graphicx}
\usepackage{tikz}
\usetikzlibrary{automata, positioning, arrows}
\usepackage{graphicx}
\usepackage[numbers]{natbib}
\usepackage{float}
\usepackage{float}
\usepackage{amsmath,amssymb,amsfonts}
\usepackage{algorithmic}
\usepackage{graphicx}
\usepackage{textcomp}
\usepackage{hyperref}
\usepackage{xcolor}
\usepackage{tikz}
\usepackage{caption}
\usepackage{subcaption}
\usepackage{longtable}
\usepackage[toc,page]{appendix}
\begin{document}

\title{Predicting Stability of Community Members in Complex Networks}

\address{sruthyksreedharan@gmail.com}

\author{Sruthi K S\thanks{sruthyksreedharan@gmail.com}\\
Department of Computer Applications\\
Cochin University of Science and Technology \\ Kochi, India\\
\and Divya Sindhu Lekha\thanks{divi.lekha@gmail.com}\\
Indian Institute of Information Technology \\
Kottayam\\
Kerala, India\\
\and A Sreekumar\thanks{sreekumar@cusat.ac.in}\\
Department of Computer Applications\\
Cochin University of Science and Technology \\ Kochi, India\\
\and Kannan Balakrishnan \thanks{mullayilkannan@gmail.com}\\
Department of Computer Applications\\
Cochin University of Science and Technology \\ Kochi, India
} 

\maketitle

\runninghead{K S Sruthi, D S Lekha, A Sreekumar and K Balakrishnan}{Stability of Community Members - Complex Networks}

\begin{abstract}
  In this work, we analyse and predict the stability of communities in complex networks. We use a variant of closeness centrality, known as profile closeness, to measure the loyalty of a member towards its community. We show that  the profile closeness is an adequate indicator of how communities evolve in a network. We investigate this in static as well as dynamic (temporal) networks and establish the relevance of profile closeness in predicting the evolution of a complex network.
\end{abstract}


\section{Introduction}
\label{sec:introduction}
    The most promising characteristic of a real complex network is the small world nature it exhibits irrespective of its size ~\cite{albert2000error,barabasi1999emergence,albert1999diameter}. A small-world network has minimal characteristic path length, owing to its significant local clustering capabilities~\cite{watts1998collective}.  In such networks, nodes tend to form densely connected regions or communities~\cite{girvan2002community}. While the nodes exhibit high connectivity within their community, they have very few connections to reach out to the nodes in other communities. Robustness of a community depends on the extent of interactions between its members. High level of intra-community interactions and existence of inter-community relations lead to small path lengths. Thus, the relative importance of a community member depends on their influence on other community members and the network as a whole. 
    
    Guimer\'{a} and Amaral (2005)~\cite{guimera2005functional} studied the pattern of intra-community connections in metabolic networks. They analyzed the degrees of nodes within the community (within-module degree) to understand if it is centralized or decentralized. A community is centralized if its members have different within-module degrees.
    
    Wang et al. (2011)~\cite{wang2011identifying} proposed two kinds of significant nodes in communities:  \emph{community cores and bridges}. Community cores are the most central nodes within the community, whereas bridges act as connectors between communities. Han et al. (2004)~\cite{han2004evidence} has also given a similar characterization of nodes important in a community as \emph{party hubs and date hubs} where party hubs are like community cores, and date hubs are like bridges. 
    
    Br\'{o}dka et al. (2012)~\cite{brodka2012predicting} experimented on evolving social networks and observed that the community prediction based on simple features are highly accurate, and that the parameters used can highly influence community prediction.
    
    Takaffoli et al. (2014)~\cite{takaffoli2014community} established that community evolution can be predicted accurately, but the predictability depends on the different events and transitions.
    
    Du et al. (2015)~\cite{du2015tracking} proposed a framework based on optimization to predict the evolution of community strength.
    
    Gupta et al. (2016)~\cite{gupta2016centrality} proposed a community-based centrality known as \emph{Comm Centrality} to find the influential nodes in a network. The computation of this centrality does not require the entire global information about the network, but only the intra- and inter-community links of a node.
    
    Recently in 2020, Zhao et al.~\cite{zhao2020identifying} proposed a method for finding the influential nodes that combines node closeness and community structure. 
    
    Kuppevelt et al.(2020)~\cite{kuppevelt2022community} put forward a metric known as community membership consistency for identifying the community core and also a node-level perception of community membership.
    
    The above works indicate that the communities, especially the relative importance of their members, influence the overall behaviour of the network considerably.

In this study, we attempt to predict the stability of a community based on profile closeness, a variant of closeness centrality. Profile closeness was introduced in \cite{lekha2020central} as a measure to detect secondary targets in a backup attack plan. We use the same concept here to analyze the fragility of members in a community. 

    \section{Computing profile closeness}
    Consider a large network $N$ with $n$ nodes and $m$ links. Then a profile is a weighted subset of nodes. 
    $$\pi = \{(u, r(u))\}$$
    where $u$ is an arbitrary vertex of $N$ and $r(u)$ is the rank of $u$ in $\pi$ based on its priority. 
    
    $N$ may contain disconnected components. When two nodes are unconnected, the distance between them becomes infinity. 
    Given a node $v$, the total distance of $v$ with respect to $\pi$ is 
    $$D_{\pi}(v) = \sum_{u \in \pi, u \neq v}d(u,v)\times r(u)$$
    Note that if $v$ and any $u \in \pi$ are in disconnected components, then $D_{\pi}(v)$ will be $\infty.$
    
    Now, we define the profile closeness $c_{\pi}(v)$ as the normalized inverse of $D_{\pi}(v)$. 
    $$c_{\pi}(v)= \frac{n}{D_{\pi}(v)}.$$
    When $v$ and any node in $\pi$ are disconnected, $c_{\pi}$ becomes zero.
    As in the case of a normal closeness centrality, nodes with higher $c_{\pi}$ values are the ones with better access to profile nodes. 
    
    \subsection{Choosing rank function}
    
    Degree ($\delta$) of a node refers to the number of edges incident on it. A high-degree node has a direct influence on a larger part of the network (See Opsahl et al.~\cite{opsahl2010node}). Therefore, it is a potentially important decision-maker in the consensus problem. Such nodes should be given a higher priority. We can do this by assigning $r(u) \rightarrow \delta(u)$.

    However, the choice of the rank function depends on the problem. An excellent candidate for the rank function in issues regarding spreading dynamics, such as information (rumour) dissemination or epidemic outbreak, is the node influence. An example of this can be the  \emph{epidemic impact} discussed in~\cite{vsikic2013epidemic}.
    
    \subsection{Choosing a profile}
    The relevance of a profile depends on the proportion of high-rank nodes included in it. If $\pi$ consists of prominent nodes (say, hubs) from different disconnected components in $N$, then it follows that $c_{\pi}$ effectively captures the relative closeness of a node to the critical nodes in $N$. A high $c_{\pi}(v)$ indicates that $v$ can act as a crucial access point to the vital areas of the network.  There are several ways to identify a set of critical nodes in a network. Refer~\cite{lu2016vital} for a state-of-art review of critical-node identification. 
    
    Detecting a set of vital nodes can help adopt budget-constrained methods to enhance the security of a network. But, this does not hold when the identified set itself is very large. In such a case, we need to find the minimum number of nodes which have easy access to this set. We can use profile closeness to evaluate this accessibility. We denote the set of vital nodes as the profile $\pi$, rank the nodes based on their vitality, compute $c_{\pi}$, and identify nodes with higher $c_{\pi}$ values. Let $k$ be the maximum number of nodes that can be secured within the given budget. Then, $k$ nodes with highest possible $c_{\pi}$ values are the efficient candidates that ought to be protected.
    
    \section{Closeness and profile closeness}
    As discussed in the introduction, the profile closeness of a node $v$ measures its closeness centrality when the profile is the entire node set and rank of the nodes is unity. i.e. 
    $$c_{\pi}(v) = c_C(v)$$
    when $\pi=V(N) \times \{1\}$.
    
    In 1979, Freeman~\cite{freeman1978centrality} introduced the concept of centralization of a graph or network to compare the relative importance of its nodes. Centralization is also a way to compare different graphs based on their respective centrality scores.

    In order to find the centralization scores, we need to determine the maximum possible value of centrality ($c_{\pi}^*$) and the deviation of the centrality of different nodes ($c_{\pi}(v)$) from $c_{\pi}^*$. The centralization index $C_{c_{\pi}}$ is the ratio of this deviation to the maximum possible value for a graph containing the same number of nodes.
    
    Freeman~\cite{freeman1978centrality} showed that the closeness centrality attains the maximum score if and only if the graph is a star. This was proven later by Everett et al.~\cite{everett2004some}. Also, the minimum value is attained when the graph is complete or a cycle. 
    
    The profile closeness $c_{\pi}$ attains the maximum value when $\pi$ is the entire set of the graph vertices. In this case, $c_{\pi}(v) = c_C(v)$ for any node $v$. Therefore, the centralization of the profile closeness coincides with the closeness centrality.
    
    However, we need to compare the performance of $c_C$ and $c_{\pi}$ for the intended applications of $c_{\pi}$. As $c_C$ is a global measure, whereas $c_{\pi}$ is highly localized to the profile $\pi$, the comparisons need to be done locally as well. So, two comparisons need to be done - one with the global closeness centrality $c_C$, and the other with a local closeness measure known as cluster closeness, $c_{cluster}$. Note that the only difference here is that $c_{cluster}$ does not have the priority ranking of group members, which is an essential feature of $c_{\pi}$. 
    
    We generate some random scale-free networks and identify their clusters. Subsequently, we calculate the global closeness $c_C$ for each node. We calculate the $c_{cluster}$ of a node as its closeness to its parent cluster. Besides, we construct a profile with these clusters. Here, the rank of a node $v$, $r(v)$, is $\delta_{cluster}(v)$(the number of neighbors of $v$ within the cluster). Thus, if a node has a large number of connections within its cluster, then it is considered as having higher priority in the profile. We compute $c_{\pi}$ with these profiles and compare them with $c_C$ and $c_{cluster}$ over all the generated networks. For comparing these measures, we use the correlation between them.

    \subsection*{Simulating correlation}
    We performed simulations on random scale-free networks with $50, 100, 500,$ and $1000$ nodes and average degrees $2, 5,$ and $7$. The results of the correlation are shown in tables~\ref{tab:cc_corr} and~\ref{tab:cluster_corr}. The values in each cell are the average correlation between the measures. The range of correlation (max-min) is shown below each value in brackets.
    
    \begin{table}[H]
        \centering     
        \begin{tabular}{ |c|c|c|c|c|}
            \hline
            \textbf{$\delta_{average}$} & \textbf{50}& \textbf{100}&\textbf{500}& \textbf{1000}\\
            \hline
            2 
            & 0.516 & 0.617 &0.782& 0.833\\
            & [0.864-0.124]  & [0.879-0.272] & [0.944-0.605] & [0.935-0.658] \\
            \hline
            5 
            & 0.522 & 0.628 & 0.805 & 0.857 \\
            & [0.793-0.128] & [0.816-0.247] & [0.900-0.684] & [0.924-0.710]\\
            \hline    
            7 
            & 0.480 & 0.617 & 0.817 & 0.872\\
            & [0.732-0.054] & [0.803-0.312] & [0.900-0.660] & [0.930-0.692]\\
            \hline        
        \end{tabular}
        \caption{Correlation between closeness and profile closeness}
        \label{tab:cc_corr}    
    \end{table}
    
    Table~\ref{tab:cc_corr} shows the correlation between the closeness centrality and profile closeness for the generated random networks. Both are positively correlated, and the relationship is reasonably good enough. An important point here is that the closeness centrality in large networks is highly correlated with its profile closeness. This fact seems interesting because the computation of profile closeness is less data-consuming when compared to the calculation of closeness centrality. Assume that both measures give the same ranking of nodes in a large network $N$. Then, we can use the low-computational profile closeness for the closeness ranking of nodes in $N$. However, more investigations need to be done in this regard. We need to perform the analysis of the simulation on vast network   to ensure this capability of profile closeness. 
    
    \begin{table}[H]
        \centering    
        \begin{tabular}{ |c|c|c|c|c|}
            \hline
            \textbf{$\delta_{average}$} & \textbf{50}& \textbf{100}&\textbf{500}& \textbf{1000}\\
            \hline
            2 
            & 0.953 & 0.960 &0.962& 0.980\\
            & [1.0-0.595]  & [0.997-0.734] & [0.999-0.049] & [0.999-0.923] \\
            \hline
            5 
            & 0.947 & 0.948 & 0.965 & 0.970 \\
            & [0.999-0.514] & [1.0-0.653] & [0.999-0.646] & [0.999-0.752]\\
            \hline    
            7 
            & 0.957 & 0.949 & 0.953 & 0.968\\
            & [0.999-0.748] & [1.0-0.537] & [0.999-0.595] & [1.0-0.706]\\
            \hline        
        \end{tabular}
        \caption{Correlation between cluster closeness and profile closeness}
        \label{tab:cluster_corr}    
    \end{table}
    Table~\ref{tab:cluster_corr} shows the correlation between cluster closeness and profile closeness for the generated random networks. We observed that the average correlations are high, which indicates a strong relationship between $c_{cluster}$ and $c_{\pi}$. Another interesting observation is that the average correlation increases steadily with network size for sparse as well as dense networks.
    \section{Predicting community stability}
    When the profile under consideration is a community, we call it a \emph{community profile}.  A community profile captures the relative importance of the community members. Here, all the nodes are not considered homogenous and we prioritize nodes like community cores and bridges.
    The application of a community profile is two-fold. 
    \begin{itemize}
        \item The community cores and bridges are prioritized in all the communities in a profile. Then, the profile closeness determines the accessibility of these vital nodes from every nook and corner of the network. This first application, the details of which are outside the scope of this work, provides a means to measure the global accessibility of the network.
        \item The community profile is constructed from a single community; with priority given to vital members. Then, the profile closeness predicts the new nodes who may join the community and members who may be on the verge of leaving the community. This second application, which will be discussed in detail in the next section, is associated with the local accessibility to a community.

    \end{itemize}
    \subsection{Construction of community profile}
    The first step in constructing a community profile is the identification of communities in the network. Once we have detected the communities, we need to rank the members in each community. The ranking is based on the intra-modular degree ($\delta_{comm}$). We can also use other relevant community-based measures like \emph{Comm centrality} (~\cite{gupta2016centrality}) for ranking. $r(v)$ denotes the rank of a node $v$. Now, we define the community profile $\pi$ as
    $$\pi = \{(v, r(v))\}$$
    The construction of a community profile is devised in algorithm~\ref{algo: comm}, Gen\_$\pi$.
    \begin{algorithm}[h]  
        \begin{algorithmic}[1]  
            \REQUIRE Community $comm=(V_{comm},E_{comm})$
            \ENSURE$\pi$
            \FOR{$v \in V_{comm}$}
            \STATE $\delta_{comm}(v) = |Neighbor_{comm}(v)|$
            \STATE $\pi \leftarrow \pi \bigcup \{(v, \delta_{comm}(v))\}$
            \ENDFOR
            \STATE  return $\pi$
        \end{algorithmic}  
        \caption{Gen\_$\pi$: Constructing community profile}  
        \label{algo: comm}  
    \end{algorithm}

    \subsection{Computing community closeness}
    Algorithm~\ref{algo: close} computes the community closeness of the entire network $cc_{\pi}[.]$     
    \begin{algorithm}[h]  
        \begin{algorithmic}[1]  
            \REQUIRE Network $N$, profile $\pi$
            \ENSURE $c_{\pi}[.]$
            \FOR{$u\in V(N)$}
            \STATE $du[.] \leftarrow SSSP(N,u)$ 
            \COMMENT {SSSP - Single Source Shortest Path}
            \STATE $D_{\pi}(u) = 0$
            \FOR{$(v,r(v)) \in \pi$}
            \IF{$(u \neq v\ and\ du[v] \neq \infty)$}
            \STATE $D_{\pi}(u) = D_{\pi}(u) + [du[v] \times r(v)]$
            \ENDIF
            \ENDFOR
            \STATE  $c_{\pi}[u] \leftarrow \frac{n}{D_{\pi}[u]}$
            \ENDFOR
            \STATE  return $c_{\pi}[.]$
        \end{algorithmic}  
        \caption{CC: Finding community closeness}  
        \label{algo: close}  
    \end{algorithm} 

    \subsection{Predicting community members}
    
    Given a node $u$ and profile $\pi$ in $N$, algorithm~\ref{algo: close} correctly computes the closeness of the node to the community corresponding to $\pi$. A community is stable when every node in a community has comparable closeness values. In other words, the community is unstable when the intra-community closeness of its nodes show drastic variations. Nodes with higher values are likely to continue in the community, whereas those with minimal values may leave the community in the future. We conducted experiments on networks with first-hand information on their ground-truth communities. Empirical evidence shows that the above observation is correct. Another interesting observation was that the nodes that exhibit more closeness towards an external community tend to join that community in future. Thus, profile closeness is an adequate indicator of how communities evolve in a network. The efficiency of this prediction depends on the design of the community profile.
    
    \subsection{Empirical evidence - Networks with ground-truth communities}
    Research on community detection has been very active for the past two decades. Many community detection techniques were devised. The Girvan-Newman method of community detection~\cite{girvan2002community}, based on edge betweenness, was a novel approach. Later, the same team came up with the \emph{modularity} concept, a qualitative attribute of a community. See ~\cite{newman2004finding}. Modularity is defined as the difference between the fraction of the edges in a community and the expected fraction in a random network. Girvan and Newman observed that, for a robust community, this attribute falls between $0.3$ and $0.7$.
    Therefore, modularity optimization can lead to better community detection. However, this is an NP-complete problem~\cite{brandes2007modularity}. Different approximation techniques based on modularity optimization produce community structures of high quality, that too with little time requirements of the order of network size.  A very recent survey by Zhao et al.~\cite{zhao2018comparative} gives a clear picture of the state-of-art in this regard.

    In this study, we used the Louvain method~\cite{blondel2008fast} of modularity optimization for detecting communities. It is an agglomerative technique with each node initially assigned as a unique community. The algorithm works in multiple passes until the best partitions are achieved. Each pass consists of two phases; in phase $1$ the nodes are moved to the neighbouring community if it can make a higher gain in modularity and in stage $2$ a new network is created from the communities detected in pass $1$.
    
    First, we simulated our results using two real-world networks in which the community structure is evident. The networks are Zachary's karate club network~\cite{zachary1977information} and the American college football network~\cite{girvan2002community}. See table~\ref{tab:Nw}.
    \begin{table}[h]
        \centering    
        \begin{tabular}{ |c|c|c|c|c|}
            \hline
            \textbf{Network} & \textbf{Nodes}& \textbf{Edges}&\textbf{Communities}& \textbf{Density}\\
            \hline
            \hline
            Karate Club & $34$ & $78$ & $2$ &$0.2781$ \\
            \hline
            College Football & $115$ & $613$ & $12$ & $0.0935164$ \\
            \hline    
            Dolphin & $62$ & $159$ & $2$ & $0.0840825$ \\
            \hline        
        \end{tabular}
        \caption{Networks with ground-truth communities}
        \label{tab:Nw}    
    \end{table}
    \subsubsection{Zachary's karate club network}
    We conducted our primary survey on the famous karate club network data collected and studied by Zachary~\cite{zachary1977information} in 1977. In his study, Zachary closely observed the internal conflicts in a 34-member group (a university-based karate club) over a period of $3$ years. The conflicts led to a fission of the club into two groups. See table~\ref{tab:Karate}. He modeled the fission process as a network. The nodes of the network represented the club members and edges represented their interactions outside the club. Zachary predicted this fission with greater than $97\%$  accuracy and argues that his observations are applicable to any bounded social groups. Many researchers used this network as a primary testbed for their studies on community formation in complex networks.
    
    \begin{table}[h]
        \centering    
        \begin{tabular}{|c|ccccccccc| }
            \hline
            \textbf{Community} & \multicolumn{9}{c|}{\textbf{Member nodes}}\\
            \hline
            I & 1 & 2 & 3 & 4 & 5 & 6 & 7 & 8 & 9 \\
            & 11 & 12 & 13 & 14 & 17 & 18 & 20 & 22 &   \\ 
            \hline
            II & 10 & 15 & 16 & 19 & 21 & 23 & 24 & 25 & 26 \\
            & 27 & 28 & 29 & 30 & 31 & 32 & 33 & 34 &   \\ 
            \hline            
        \end{tabular}
        \caption{Ground-truth communities in Karate network}
        \label{tab:Karate}    
    \end{table}
    
\begin{figure*}
    \centering
        \includegraphics[scale=.4]{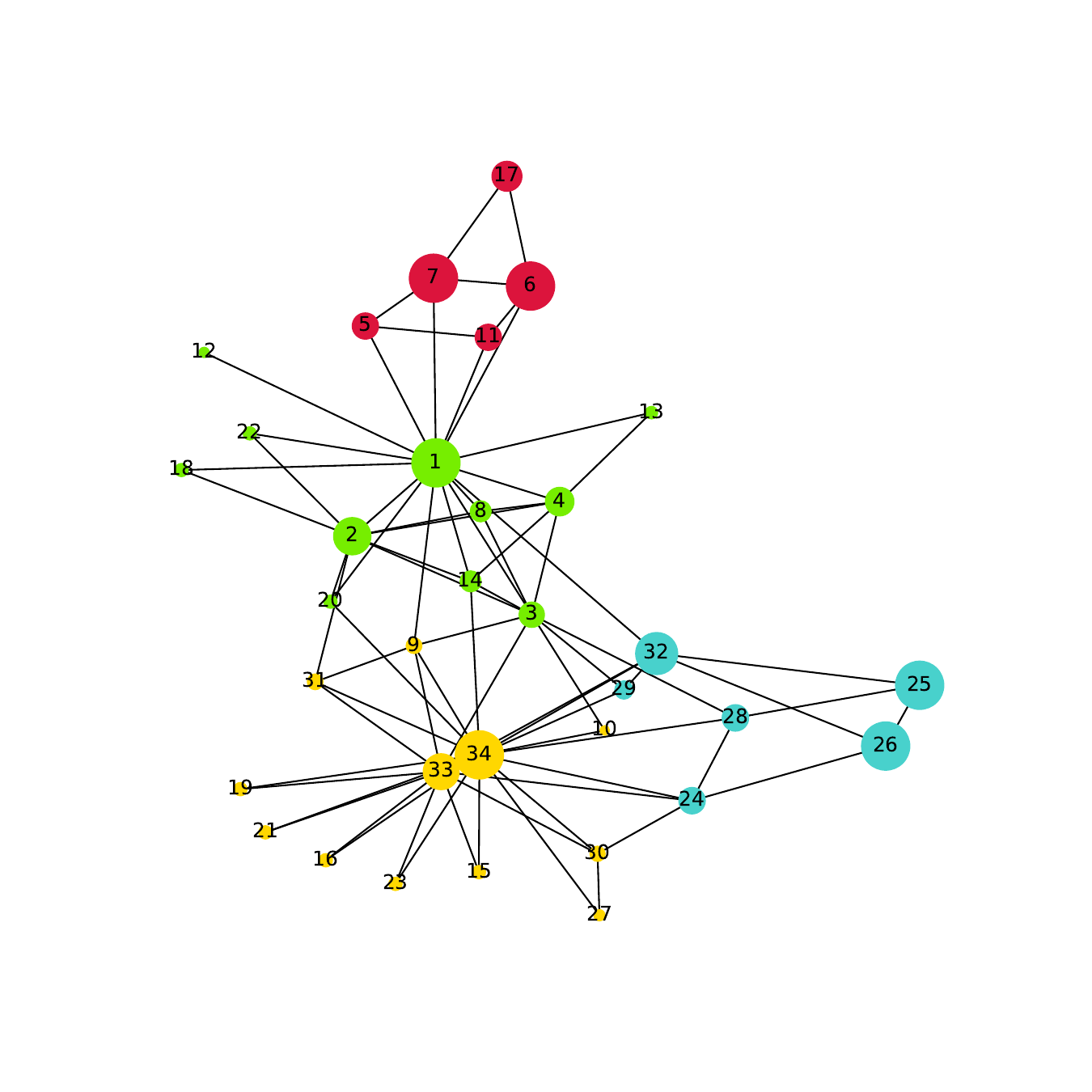}
        \caption{Community closeness in Karate club network.}
        \label{fig_karate}
    \end{figure*}

    We identified $4$ communities in the network (using the Louvain method). See table~\ref{tab:Karate1}.
    \begin{table}[h]
        \centering    
        \begin{tabular}{|c|ccccccccccc| }
            \hline
            \textbf{Comm.} & \multicolumn{11}{c|}{\textbf{Member nodes}}\\
            \hline
            I & 1 & 2 & 3 & 4 & 8 & 12 & 13 & 14 & 18 & 20 & 22 \\ 
            \hline        
            II & 5 & 6 & 7 & 11 & 17 & & & & & & \\ 
            \hline
            III & 9 & 10 & 15 & 16 & 19 & 21 & 23 & 27 & 30 & 31 & 33 \\
            & 34 & & & & & & & & & & \\ 
            \hline    
            IV & 24 & 25 & 26 & 28 & 29 & 32 & & & & &   \\     
            \hline                        
        \end{tabular}
        \caption{Communities detected in Karate network}
        \label{tab:Karate1}    
    \end{table}
    We used the intra-module degree ($\delta_{community}$) of nodes for constructing the profile. The nodes in the profile were prioritized based on their $\delta_{community}$ value. Nodes having higher value were given higher priority. Subsequently, the profile closeness was computed for each community member. See figure~\ref{fig_karate}. Different colors represent the members of different communities. The relative size of the nodes represent their profile closeness with respect to their own community.
    
    \begin{figure*}
        \centering
        \includegraphics[scale=.4]{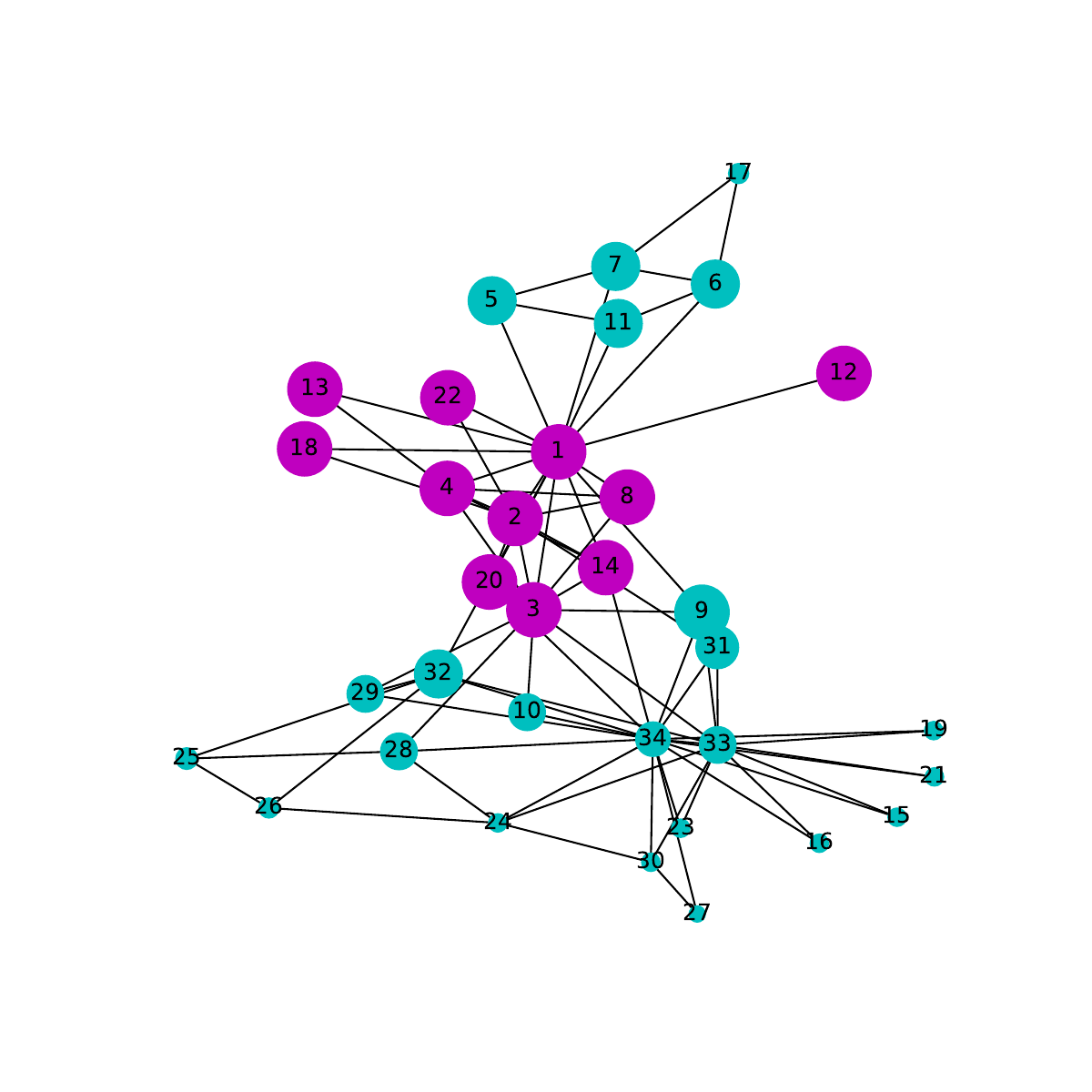}
        \caption{Profile closeness of external nodes to community $I$.}
        \label{fig_karate1}
    \end{figure*}
    
    The profile closeness of node $9$ in its community ($cc_{Community_{III}}(9)$) is very low. From this, we can interpret that $9$ has a higher tendency to leave its community. Also, we compared the profile closeness of all nodes with respect to \emph{Community I} ($cc_{Community_I}$). See figure~\ref{fig_karate1}. Nodes external to \emph{Community I} are coloured blue. Among them, \emph{Node 9} has a higher value for $cc_{Community_I}$. This high value of $cc_{Community_I}(9)$ and the low value of $cc_{Community_{III}}(9)$ indicates that $9$ has more affinity towards \emph{Community I} than its community, \emph{Community III}.
    
    This observation is relevant since node $9$ originally belonged to \emph{Community I} as noted by Zachary. Furthermore, Zachary had even observed that member $9$ is a weak supporter of the second faction ($II$); but joined the first faction ($I$) after the fission. Our method also reproduced the same fact.
    
    \subsubsection{American college football network}
    The second network chosen for our study was the American college football network, from the dataset collected by Newman~\cite{girvan2002community}. The nodes in this network represent the college football teams in the U.S., and the edges represent the games between them in the year 2000. About $8$-$12$ teams were grouped into a conference. Altogether $12$ conferences were identified. Most of the matches were between the teams belonging to the same conference. Therefore, the inherent community structure in this network corresponds to these conferences. These ground-truth communities are given in table~\ref{tab:FN}.
    
    \begin{longtable}{|c|ccc|}
    \caption{US Football Network: Ground truth communities}
            \label{tab:FN}    
        \\
                \hline
                \textbf{Conference}&\multicolumn{2}{c}{\textbf{College teams}} \\
                \hline
                \hline
                Atlantic & Flora. St. & N. Caro. St.    &  Virginia \\
                Coast & Georg. Tech &  Duke & N. Caro. \\ 
                & Clemson & Maryland &  Wake Forest\\
                \hline
                IA & Cent. Flora & Connecticut &  Navy\\
                Independents & Notre Dame &  Utah St. & \\
                \hline
                Mid & Akron & Bowl. Green St.&  Buffalo    \\
                American & Kent &  Miami Ohio & Marshall\\ 
                & Ohio & N. Illin. &  W. Michigan\\
                & Ball St. &  C. Michigan & Toledo\\
                &  E. Michigan &  &\\
                \hline            
                Big & Virg. Tech & Boston Coll. & W. Virg.\\
                East & Syracuse & Pittsburg & Temple\\ 
                & Miami Flora & Rutgers &\\
                \hline
                Conference & Alabama Birm. & E. Caro. & S. Missis.\\
                USA & Memphis & Houston & Louisville\\ 
                & Tulane & Cincinnati & Army \\
                & T. Christ. &  &\\
                \hline                    
                SEC    & Vanderbilt & Florida & Kentucky \\
                & S. Caro. & Georgia & Tennessee\\ 
                & Arkansas & Auburn & Alabama\\
                & Missis. St. & Louis. St. & Missis.\\
                \hline                    
                W. & Louis. Tech & Fresno St. & Rice\\
                Athletic & S. Method. & Nevada & San Jose St.\\ 
                & T. El Paso & Tulsa & Hawaii\\
                & Boise St. & & \\
                \hline                    
                Sun & Louis. Monroe & Louis. Lafay. & Mid. Tenn. St.\\
                Belt & N. Texas & Arkansas St. & Idaho\\ 
                & New Mex. St. & &\\
                \hline                    
                Pac & Oreg. St. & S. Calif.    & UCLA \\
                10 & Stanford & Calif. & Ariz. St.\\ 
                & Ariz. & Washing. & Washing. St.\\
                & Oregon & &\\
                \hline            
                Mountain & Brigh. Y. & New Mex.    & San Diego St.\\
                West & Wyoming & Utah & Colorado St.\\ 
                & Nev. Las Vegas & Air Force &\\
                \hline
                Big & Illin. & Nwestern & Mich. St.\\
                10 & Iowa & Penn St. & Mich.\\ 
                & Ohio St. & Wisconsin & Purdue \\
                & Indiana & Minnesota &\\
                \hline
                Big & Oklah. st.& Texas & Baylor \\
                12 & Colorado & Kansas & Iowa St.\\ 
                & Missouri & Nebraska & Texas Tech \\
                & Texas A \& M & Oklahoma & Kansas St.\\
                \hline
    \end{longtable}
    
    In the community detection step, we identified ten communities (See table~\ref{tab:FN1}). Four among them ($I$, $VII$, $IX$ and $X$) correspond to the ground-truth communities (\emph{AtlanticCoast, Pac 10, Big 10 and Big 12} respectively.) Community $VIII$ is a combination of two actual communities, \emph{Mountain West} and \emph{Sun Belt}.

\begin{figure*}    
    \includegraphics[scale=.6]{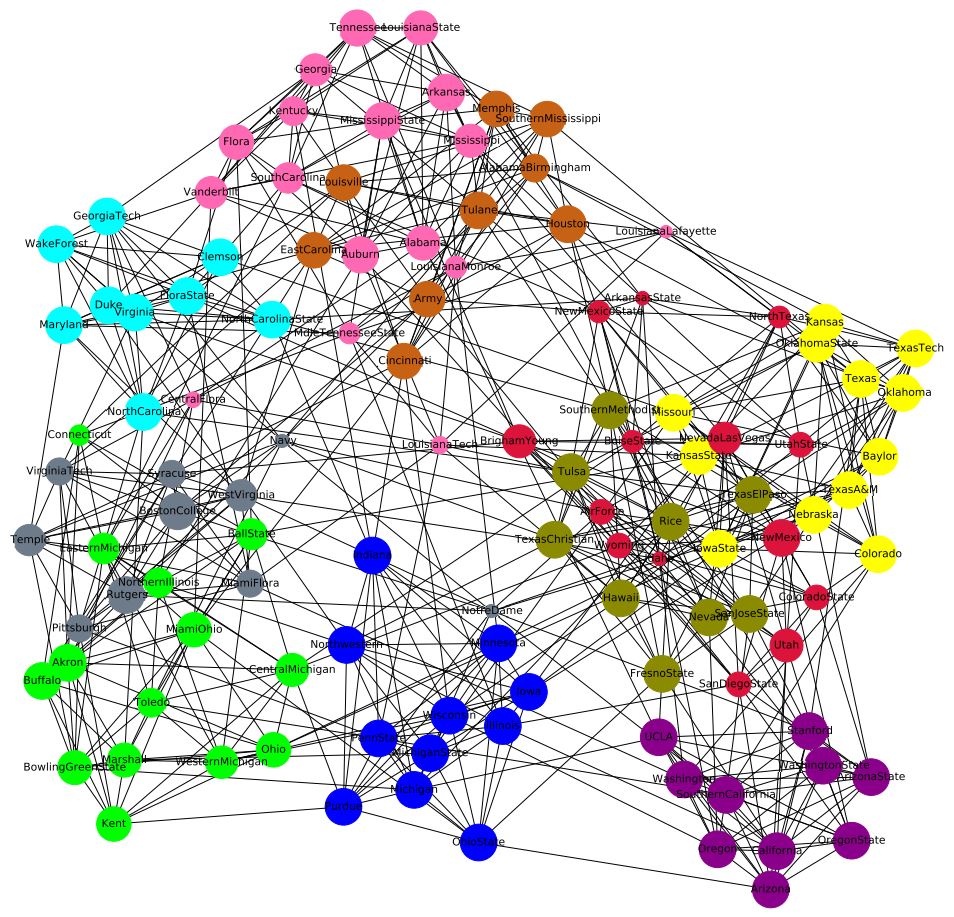}
        \caption{Community closeness in American college football network.}
        \label{fig_football}
    \end{figure*}

    \begin{longtable}{|c|ccc|}
            \caption{US Football Network: Detected communities}
        \label{tab:FN1} 
        \\
            \hline
            \textbf{Community}&\multicolumn{2}{c}{\textbf{Member teams}} \\
            \hline
            \hline
            I & Flora St. & N. Caro. St.&  Virginia \\
            & Georg. Tech &  Duke & N. Caro.\\ 
            &  Clemson & Maryland &  Wake Forest \\
            \hline
            II &  Connecticut &  Toledo  &  Akron \\
            & Bowl. Green St. &  Buffalo & Kent \\
            &  Miami Ohio & Marshall &  Ohio \\
            & N. Illin.    &  W. Mich.    & Ball St. \\
            &  C. Mich. & E. Mich. \\
            \hline            
            III & Virg. Tech & Boston Coll. & W. Virg.\\
            & Syracuse & Pittsburg & Temple\\ 
            & Miami Flora & Rutgers & Navy \\
            & Notre Dame & &\\
            \hline
            IV & Alabama Birm. & E. Caro. & S. Missis.\\
            & Memphis & Houston & Louisville\\ 
            & Tulane & Cincinnati & Army  \\
            \hline                    
            V & Vanderbilt & Flora & Kentucky \\
            & S. Caro. & Georgia & Tennessee\\ 
            & Arkansas & Auburn & Alabama \\
            & Missis. St. & Louis. St. & Missis.\\
            & Louis. Monroe & Mid. Tennes. St.    & Louis.Lafay.\\
            & Louis. Tech &  C. Flora & \\
            \hline                    
            VI & Rice & S. Method. & Nevada\\
            & San Jose St. & T. El Paso & Tulsa \\
            & Hawaii & Fresno St. & T. Christ. \\
            \hline                    
            VII & Oregon St. & S. Calif. & UCLA\\
            & Stanford & Calif. & Arizona St.\\ 
            & Arizona & Washing. & Washing. St.\\
            & Oregon & & \\
            \hline            
            VIII & Brigham Y. & New Mex. & San Diego St.\\
            & Wyoming & Utah & Colorado St.\\ 
            & N Las Vegas & Air Force & Boise St.\\
            & N. Texas & Arkansas St.& New Mex. St.\\
            & Utah St. & Idaho & \\
            \hline
            IX & Illinois & Nwestern & Mich. St. \\
            & Iowa & Penn St. & Michigan\\ 
            & Ohio St. & Wisconsin & Purdue\\
            & Indiana & Minnesota &\\
            \hline
            X & Oklah. st. & Texas & Baylor\\
            & Colorado & Kansas & Iowa St.\\ 
            & Missouri & Nebraska & Texas Tech \\
            & Texas A \& M & Oklahoma & Kansas St.\\
            \hline
    \end{longtable}
    We computed the community closeness of nodes. See figure~\ref{fig_football}.
    
\begin{figure*}
        \includegraphics[scale=.6]{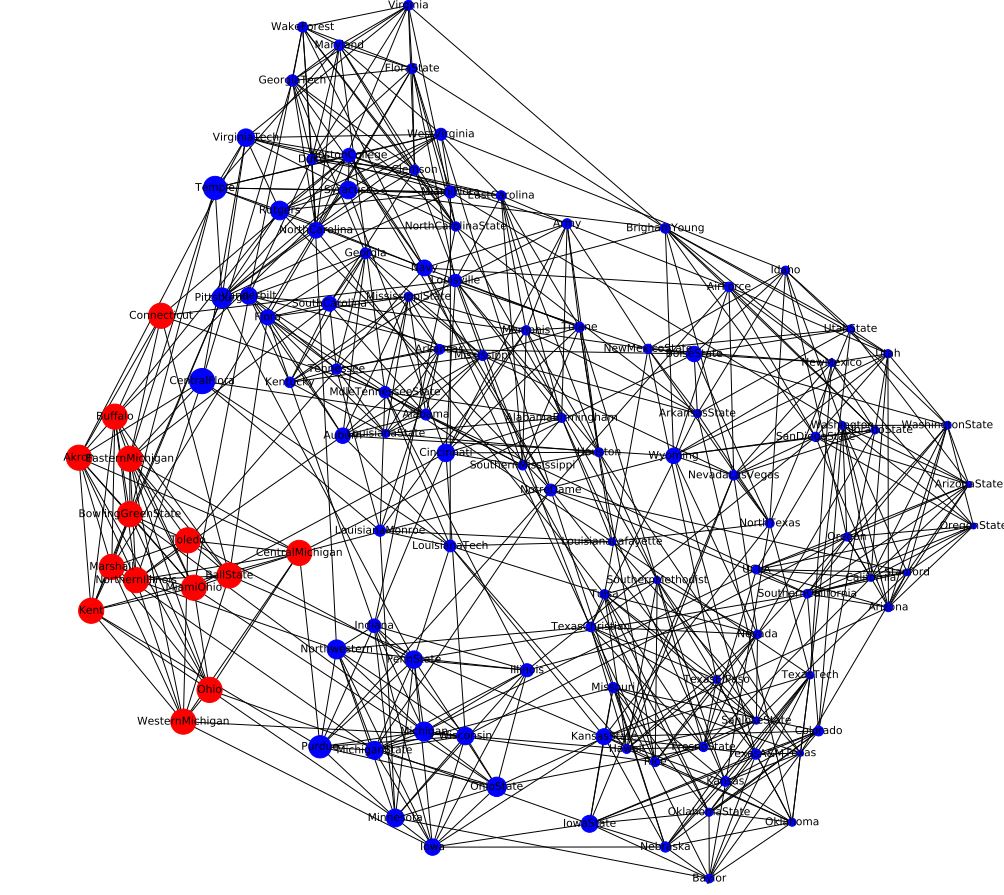}
        \caption{Profile closeness of external nodes to community $II$ of American college football network.}
        \label{fig_fb_comm1}
    \end{figure*}

    We then examined the profile closeness of all the nodes to community $II$. See figure~\ref{fig_fb_comm1}. We observed that \emph{Central Florida} has a greater closeness to $II$. This observation conforms to the ground truth that \emph{Central Florida} team played with teams like \emph{Connecticut} in many matches.

    \subsubsection{Dolphins network}
    Another chosen network with the ground-truth community is the dolphins network, which is from the dataset collected by Lusseau et al., in the University of Otago- Marine Mammal Research Group~\cite{lusseau2003bottlenose} (2003). Lusseau along with Newman~\cite{lusseau2004identifying} (2004) used this data to study social networks of bottlenose dolphins. In their study, they observed fission in the network to two groups with one individual (\emph{SN100}) temporarily leaving the place. These communities are shown in table~\ref{tab: dlp}.
    \begin{table}[H]
        \centering    
        \begin{tabular}{|c|ccccc|}
            \hline
            \textbf{Group}&\multicolumn{5}{c}{\textbf{Member dolphins}}\\
            \hline
            I& Beak & Bumper & CCL & Cross & Double\\
            & Fish & Five & Fork & Grin & Haecksel\\
            & Hook & Jonah & Kringel & MN105 & MN60\\
            & MN83 & Oscar & Patchback & PL & Scabs \\
            & Shmuddel & SMN5 & SN100 & SN4 & SN63\\
            & SN89 & SN9 & SN96 & Stripes & Thumper\\
            & Topless & TR120 & TR77 & Trigger & TSN103\\
            & TSN83 & Vau & Whitetip & Zap & Zipfel \\
            \hline
            II & Beescratch & DN16 & DN21 & DN63 & Feather\\
            & Gallatin & Jet & Knit & MN23 & Mus \\
            & Notch & Number1 & Quasi & Ripplefluke & SN90\\
            & TR82 & TR88 & TR99 & Upbang & Wave \\
            & Web & Zig & & &\\
            \hline        
        \end{tabular}
        \caption{Ground-truth groups in dolphin network}
        \label{tab: dlp}
    \end{table}
    
\begin{figure*}
        \centering
        \includegraphics[scale=.6]{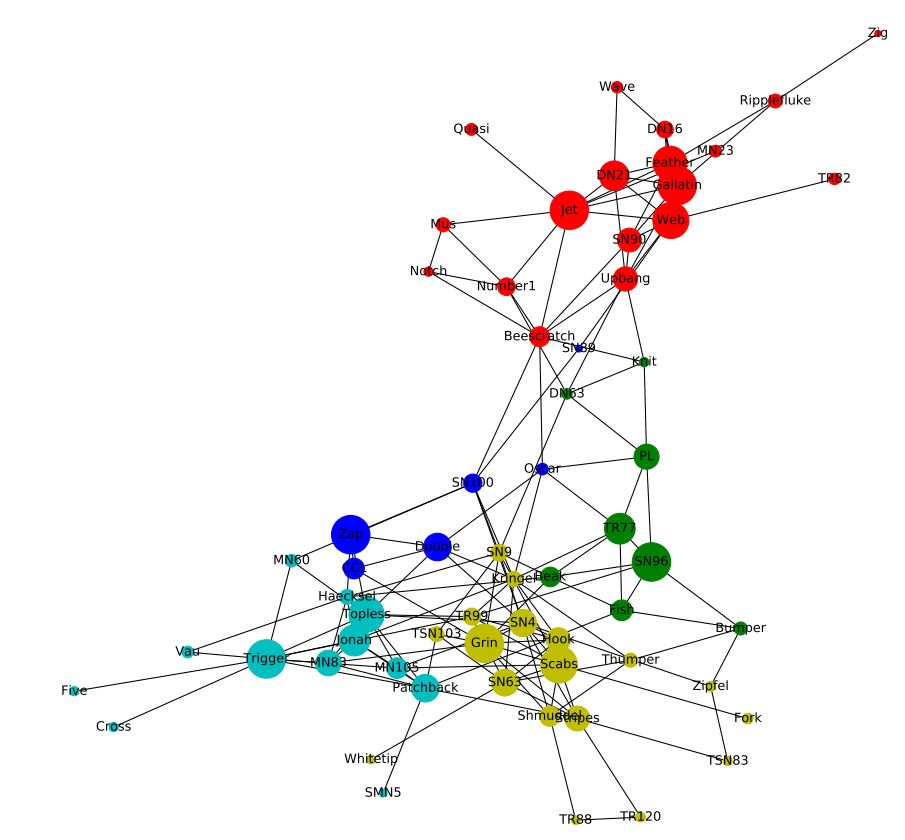}
        \caption{Community closeness in Dolphins network.}
        \label{fig_dolphin}
    \end{figure*}

\begin{figure*}
        \centering
        \includegraphics[scale=.6]{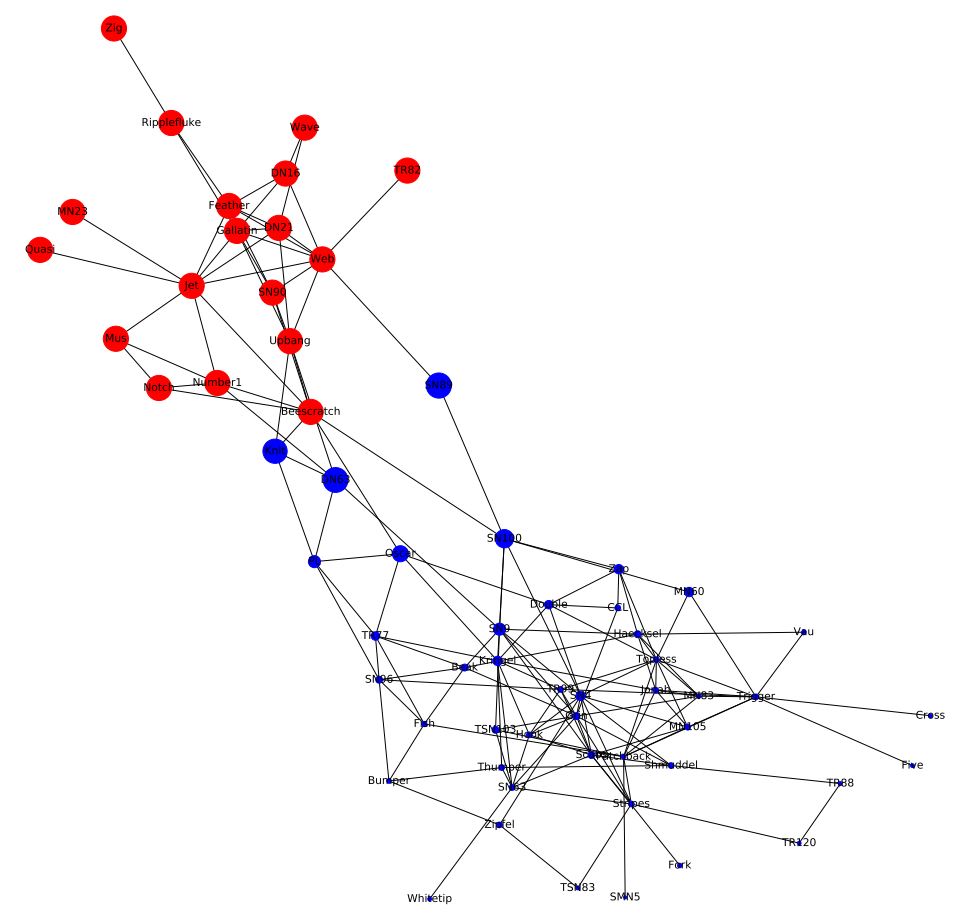}
        \caption{Community closeness in Dolphins network.}
        \label{fig_dolphin_comm}
    \end{figure*}

    We detected $5$ communities. See figure~\ref{fig_dolphin}. The communities are shown in table~\ref{tab: dlp1}.
    \begin{table}
        \centering    
        \begin{tabular}{|c|ccccc|}
            \hline
            \textbf{Group}&\multicolumn{2}{c}{\textbf{Member dolphins}} \\
            \hline
            \hline
            I & Beak & Bumper & Fish & Knit & DN63\\
            & PL & SN96 & TR77 & &\\
            \hline
            II & CCL & Double & Oscar & SN100 & SN89\\
            & Zap & & & &\\
            \hline
            III & Cross & Five & Haecksel & Jonah & MN105\\
            & MN60 & MN83 & Patchback & SMN5 & Topless\\
            & Trigger & Vau & & &\\
            \hline
            IV & Fork & Grin & Hook & Kringel & Scabs\\
            & Shmuddel & SN4 & SN63 & SN9 & Stripes\\
            & Thumper & TR120 & TSN103 & TSN83 & Whitetip\\
            & Zipfel & TR99 & TR88 & & \\
            \hline
            V & Beescratch & DN16 & DN21 & Feather & Gallatin\\
            & Jet & MN23 & Mus & Notch & Number1\\
            & Quasi & Ripplefluke & SN90 & TR82 & Upbang\\
            & Wave & Web & Zig & & \\
            \hline        
        \end{tabular}
        \caption{Communities detected in dolphin network}
        \label{tab: dlp1}
    \end{table}
    We checked the closeness to community $V$. See figure~\ref{fig_dolphin_comm}. It is clearly visible that \emph{DN63} and \emph{Knit} are having higher chances of grouping with community $V$. This conforms to the observation made by Lusseau and Newman.
    
    \section{Empirical evidence in temporal networks}
We experimented on seven dynamic real-world networks as well. The following temporal networks were chosen for analysis.
  \begin{table}
        \centering    
        \begin{tabular}{ |p{3.5cm}|p{1.5cm}|p{1.5cm}|p{1.5cm}|p{5cm}|}
            \hline
            \textbf{Network} & \textbf{Average Nodes}& \textbf{Average Edges}&\textbf{Time stamps}& \textbf{Description}\\
            \hline
            \hline
            Dutch school friendships (2003)~\cite{snijders2010introduction} & $26.5$ & $115.25$ & $4$ &Friendships between the students at secondary school in The Netherlands. \\
            \hline
            Jakarta terrorists (2009)~\cite{Atran09} & $28$ & $25.75$ & $4$ & Connections between the individuals associated with  Jakarta bombings. \\
            \hline
            Australian Embassy bombing (2004)~\cite{Atran09a} & $27$ & $41.07$ & $15$ & Connections between the individuals associated with Australian Embassy bombing in Indonesia. \\
            \hline    
            Enron email network (2011)~\cite{boldi2011layered} & $788.75$ & $2702.08$ & $12$ & Email communication between Enron email IDs  \\
            \hline    
            Autonomous systems AS-733 network (2001)~\cite{leskovec2005graphs}  & $6086.36$ & $11862.21$ & $14$ & BGP traffic among autonomous systems (ASs) on the Internet, from the Oregon Route Views Project  \\ 
            \hline    
           College messaging network (2009)~\cite{panzarasa2009patterns} & $703.83$ & $2604.67$ & $6$ & Private messages sent on an online social network at the University of California, Irvine  \\     
            \hline    
           Oregon 2 Autonomous systems network (2001)~\cite{leskovec2005graphs} & $11150$ & $31559$ & $9$ &  AS peering information inferred from Oregon route-views, Looking glass data, and Routing registry, all combined.  \\            
            \hline        
        \end{tabular}
        \caption{Temporal networks with ground-truth communities}
        \label{tab:Temp_Nw}    
    \end{table}
    
    We analysed how the evolution of communities in these networks are effected by the profile closeness of its members. Table~\ref{tab:Temp_Nw} shows results of these experiments. Almost $50\%$ of the members removed from a community are those with low intra-community closeness. But, this was not true in the case of \textit{Australian Embassy bombing network} which was a small network of $27$ nodes. However, when we analysed the node additions to communities; we could observe that the number of high value nodes added to the communities are very varying. For larger networks (AS-733, College Messaging and Oregon 2 networks), the probability was higher than that of smaller networks. We ignored the $splitting/merging$ event of community evolution here since community splitting and changing the membership of a small fraction of nodes are different scenarios. We are considering this as an immediate follow-up to this work.

    \begin{table}
        \centering    
        \begin{tabular}{ |p{6cm}|p{3.5cm}|p{4cm}|}
            \hline
            \textbf{Network} & \textbf{\% nodes with high community closeness added}& \textbf{\% nodes with low intra-community closeness removed}\\
            \hline
            \hline
            Dutch school friendships & $35.10\%$ & $52.20\%$  \\
            \hline
            Jakarta terrorists  & $4\%$ & $50\%$  \\
            \hline
            Australian Embassy bombing  & $5\%$ & $19.30\%$  \\
            \hline    
            Enron email network  & $12.90\%$ & $53.25\%$ \\
            \hline    
            Autonomous systems AS-733 network & $47.66\%$ & $43.16\%$  \\ 
            \hline    
           College messaging network & $25.62\%$ & $51.47\%$  \\     
            \hline    
           Oregon 2 Autonomous systems network & $49.50\%$ & $50.45\%$  \\            
            \hline        
        \end{tabular}
        \caption{Analysis of community evolution in temporal networks based on community closeness}
        \label{tab:Temp_Nw}    
    \end{table}
    
    Visualizations of community evolution based on community closeness is given in the Appendix. See figures~\ref{fig:dutch},~\ref{fig:jakarta}, and ~\ref{fig:australian} for the smaller networks viz. Dutch school friendship, Jakarta Terrorists, and Australian Embassy Bombing networks respectively.

    
\section{Summary}

The most noteworthy finding of this work is that the relative importance of the community members plays a crucial role in attracting new nodes or repelling existing nodes. This finding can be useful in assessing the dynamically changing degree of participation of a node to different communities.

Here, we used the concept of profile closeness centrality for predicting the stability of network communities. 

Some of the salient features of profile closeness are the following. 
\begin{itemize}
    \item The rank assigned to a profile node depends on the extent of the influence that it has on the network. For example, high degree nodes, which directly influence a large part of the network, are ranked high.
    \item Choice of the rank function depends on the domain of the network.
    \item It is suitable for budget-constrained network problems.
    \item It closely correlates with the global closeness centrality for large networks. Therefore, profile closeness offers a low computational complexity approximation of closeness ranking.
    \item It can aid in predicting community evolution.
\end{itemize}

We have also analysed the role of community closeness in understanding the repulsion of a node to its community and its affinity towards another community in temporal networks. We have concluded that on an average, half of the nodes removed from a community are those with low intra-community closeness. Also, in very large networks, half of the new nodes added to a community are those having high community closeness.

However, we need more investigations to develop alternative techniques to assign member priorities.

 \section*{Acknowledgments}
    The authors are grateful to Prof. Animesh Mukherjee (IIT Kharagpur) for providing valuable comments on a previous version of this work.
    
    This work was partially supported by the post doctoral fellowship (to Corresponding author) from Cochin University of Science and Technology, 2019-2020.

\begin{small}
\begin{verbatim}

\end{verbatim}
\end{small}

\nocite{*}
\bibliographystyle{fundam}
\bibliography{citations}

\newpage
\begin{appendices}

 \begin{figure}[h]
     \centering
     \begin{subfigure}[b]{0.4\textwidth}
         \centering
         \includegraphics[scale=0.16]{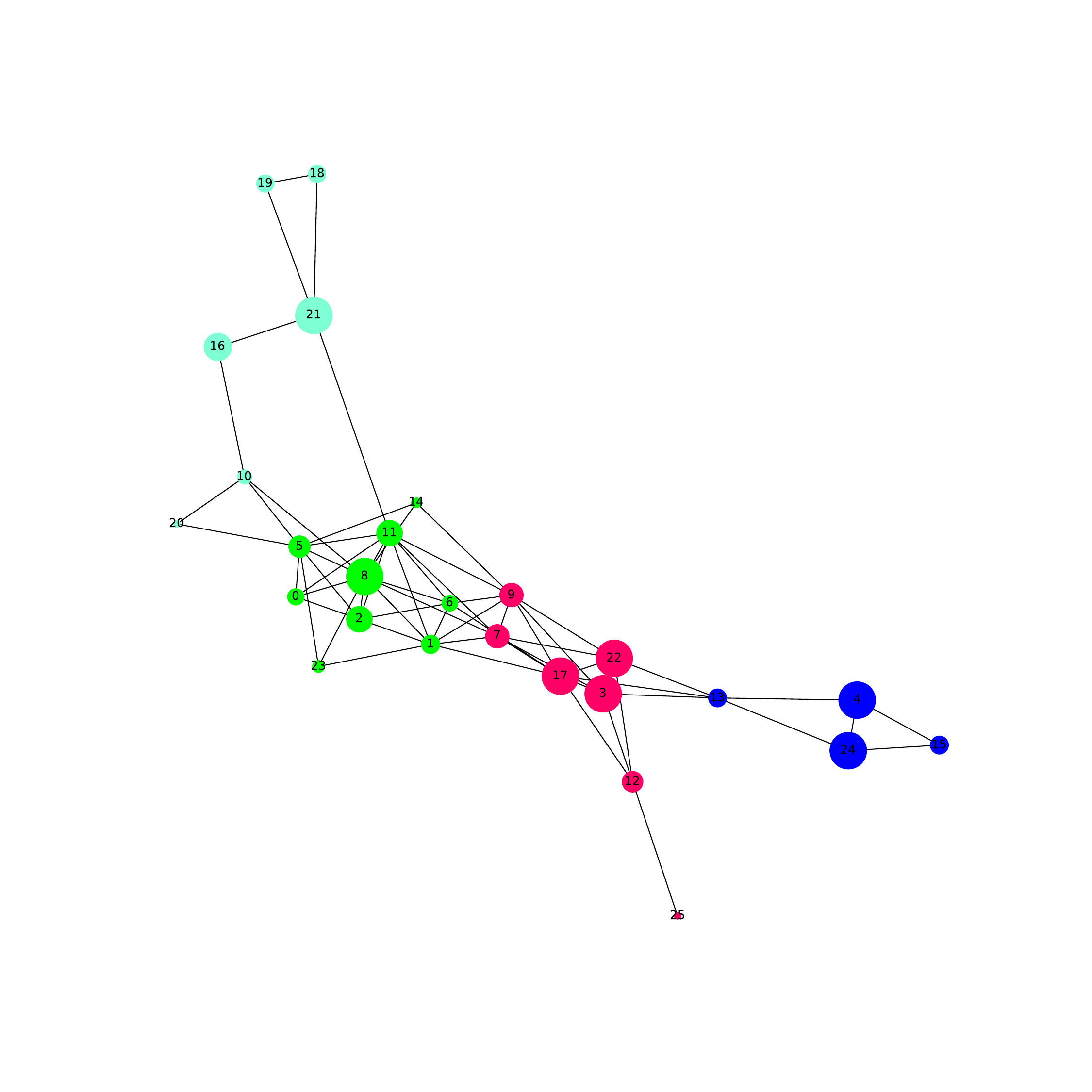}
         \caption{snapshot 1 : nodes = 26 , edges = 63, number of communities = 4 }
         \label{fig:jakarta_1}
     \end{subfigure}
     \hfill
     \begin{subfigure}[b]{0.4\textwidth}
         \centering
         \includegraphics[scale=0.16]{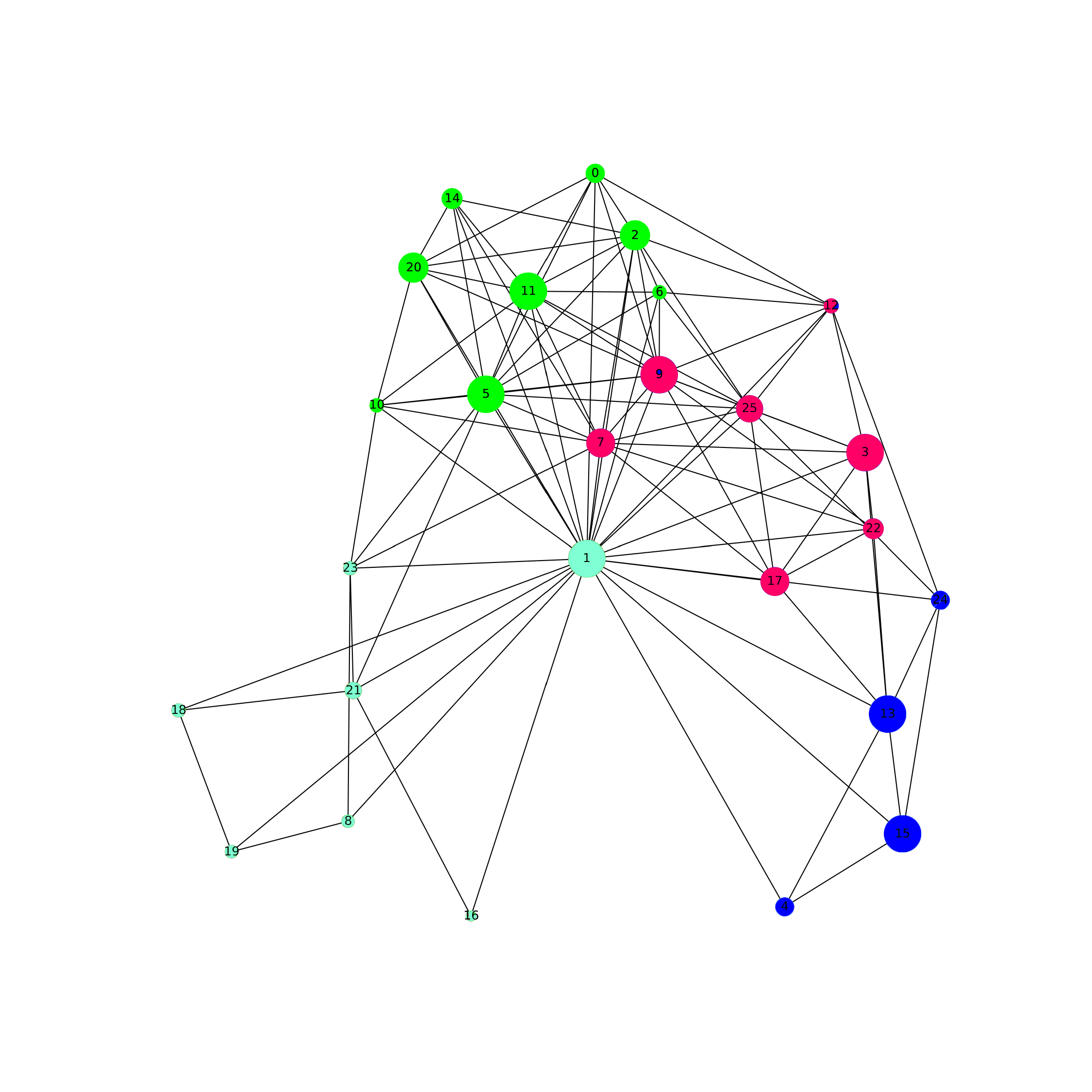}
         \caption{snapshot 2 : nodes = 26, edges = 103 , number of communities = 4}
         \label{fig:jakarta_2}
     \end{subfigure}
    \hfill
    \begin{subfigure}[b]{0.4\textwidth}
         \centering
         \includegraphics[scale=0.16]{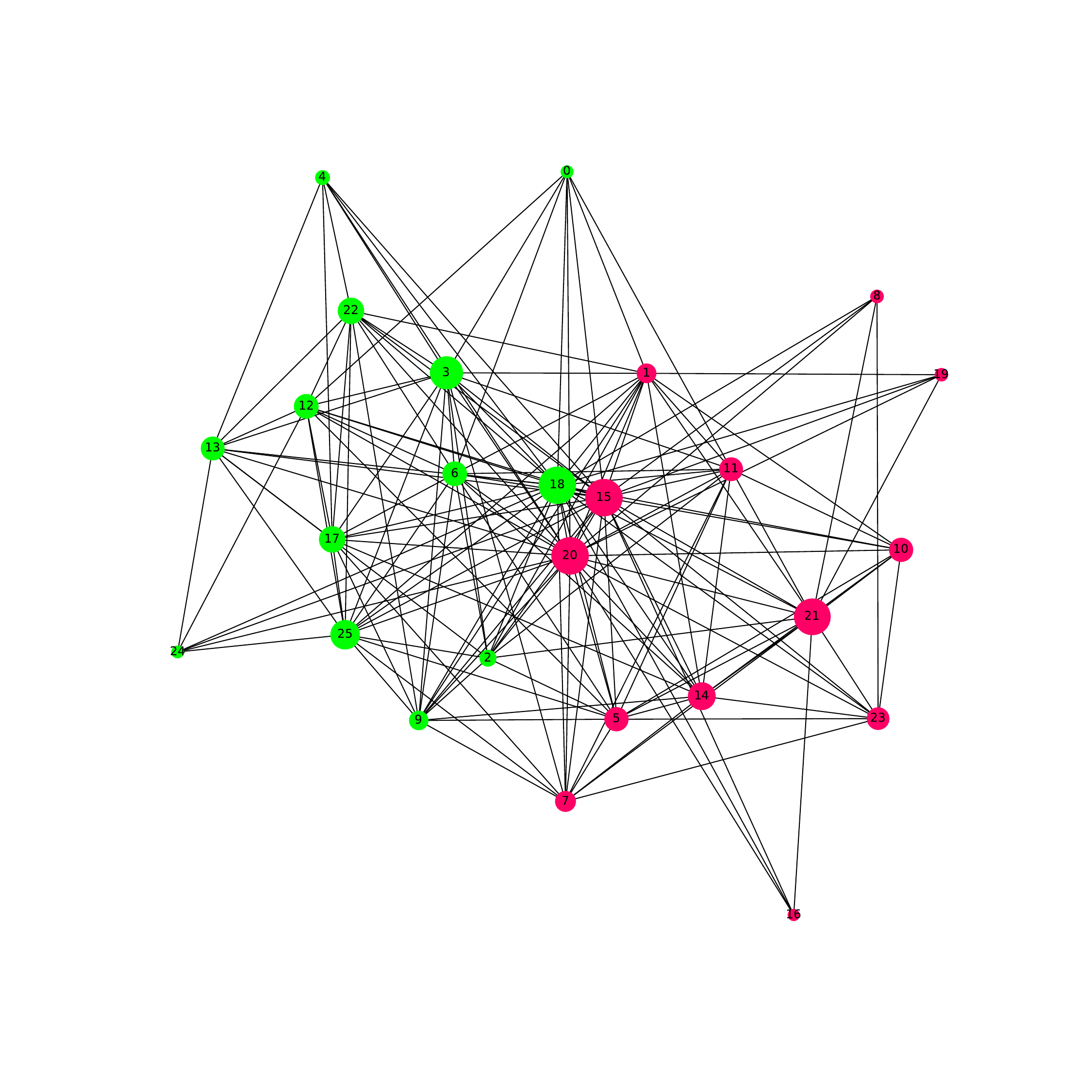}
         \caption{snapshot 3 : nodes = 26, edges = 167, number of communities = 2 }
         \label{fig:jakarta_3}
     \end{subfigure}
     \hfill
     \begin{subfigure}[b]{0.4\textwidth}
         \centering
         \includegraphics[scale=0.16]{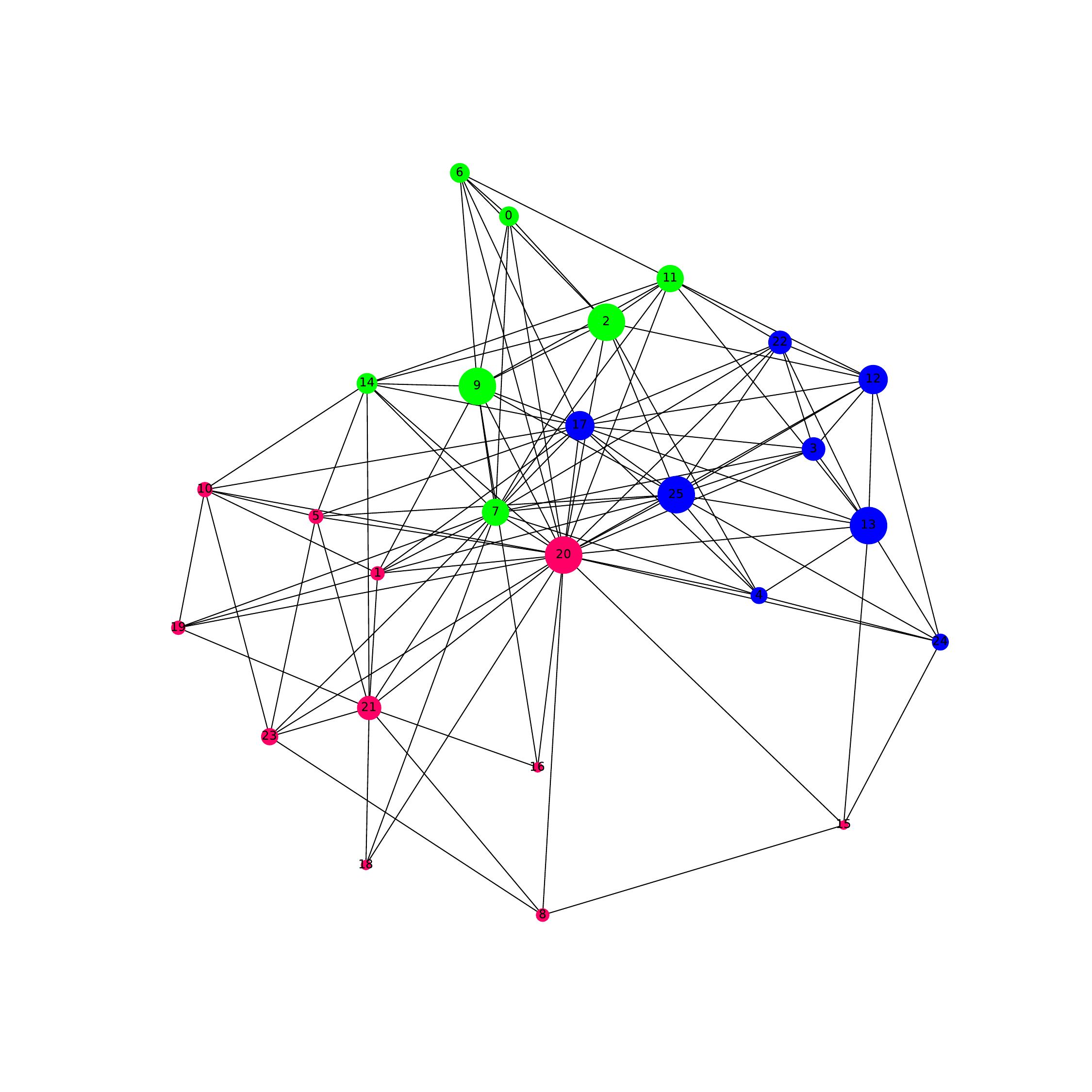}
         \caption{snapshot 4 : nodes = 26, edges = 112, number of communities = 3}
         \label{fig:jakarta_4}
     \end{subfigure}
     \hfill
    
        \caption{Community closeness in dutch school friendship network}
        \label{fig:dutch}
\end{figure}     

 \newpage
 
 \begin{figure}[h]
     \centering
     \begin{subfigure}[b]{0.4\textwidth}
         \centering
         \includegraphics[scale=0.2]{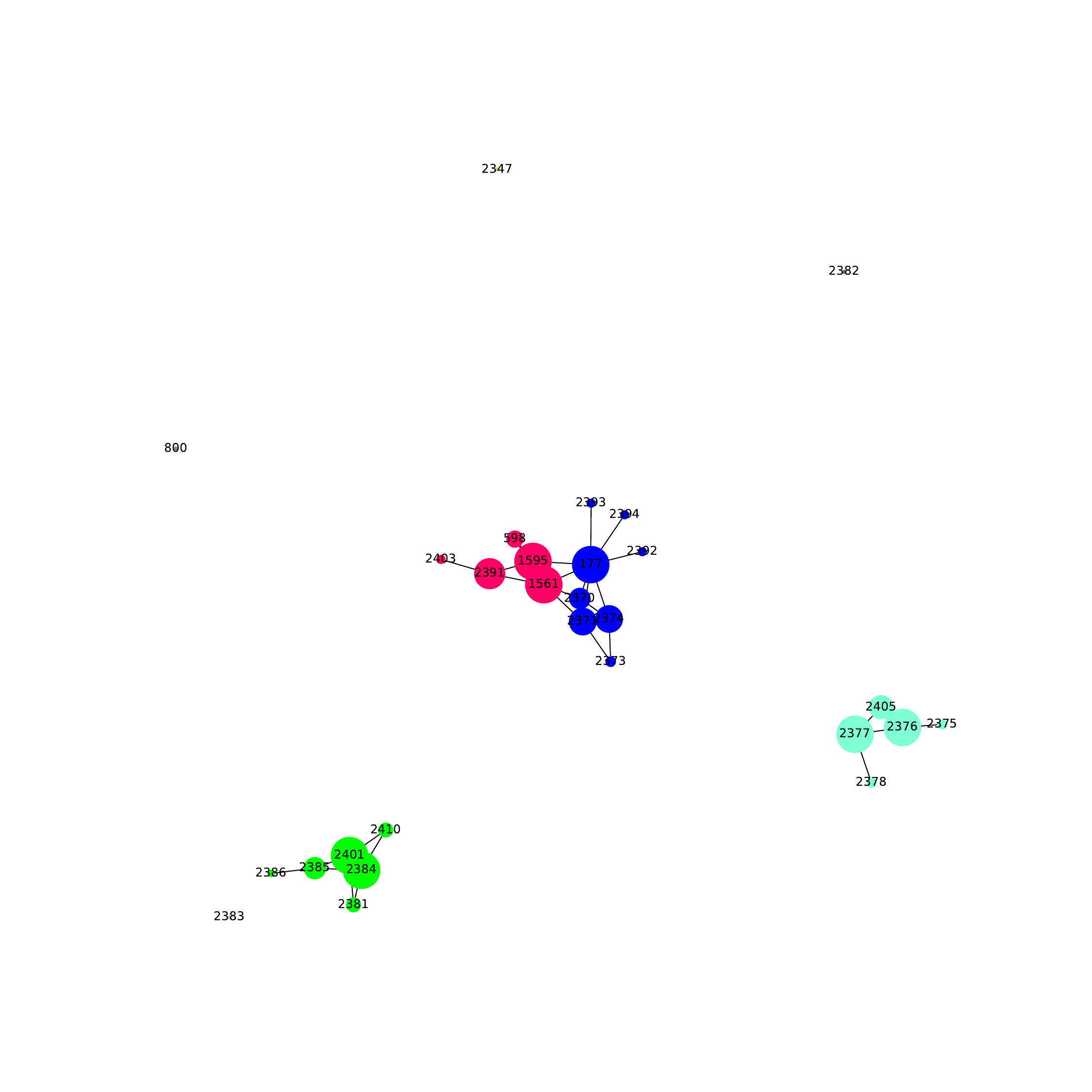}
         \caption{snapshot 1 : nodes = 28, edges = 34, number of communities = 8}
         \label{fig:jakarta_1}
     \end{subfigure}
     \hfill
     \begin{subfigure}[b]{0.4\textwidth}
         \centering
         \includegraphics[scale=0.2]{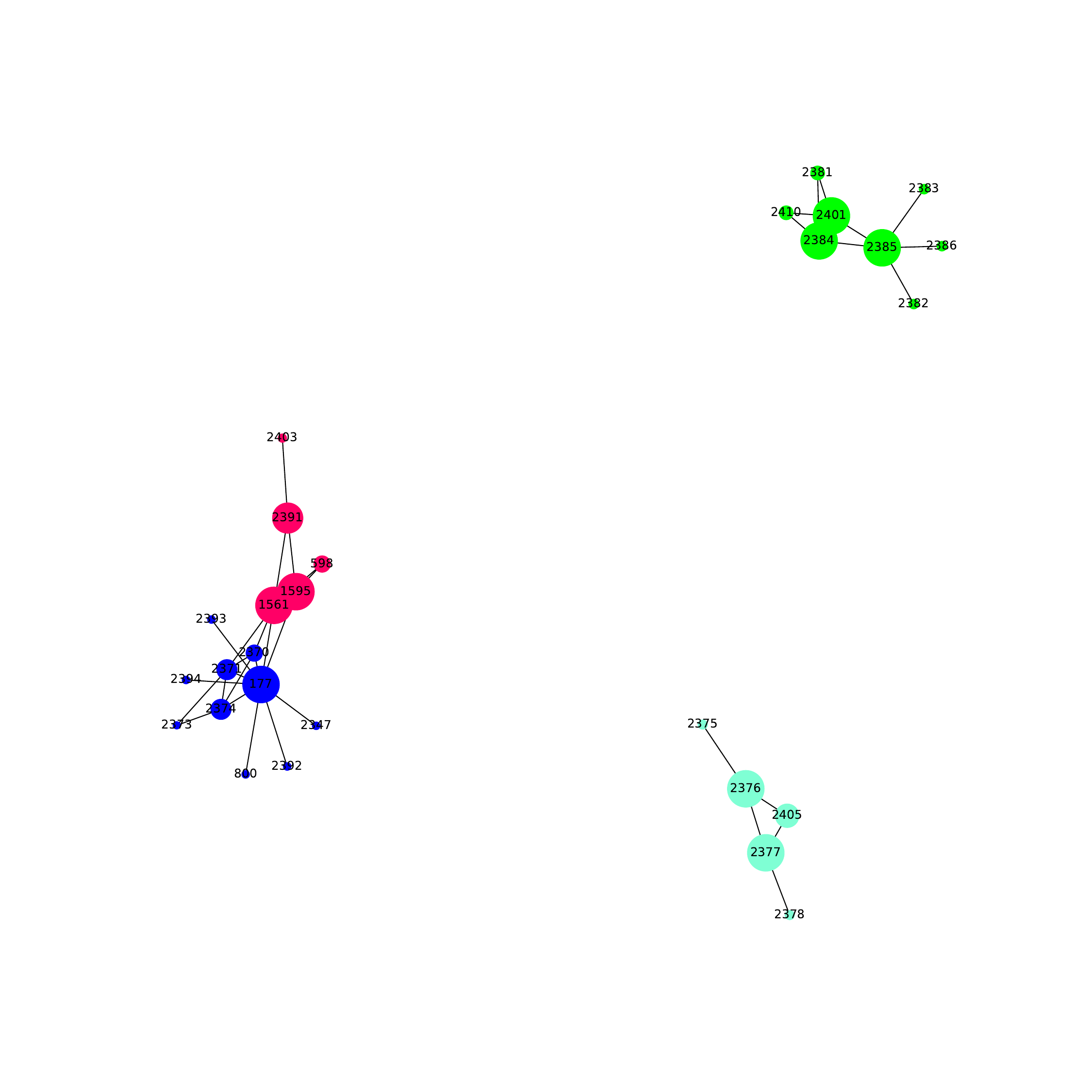}
         \caption{snapshot 2 : nodes = 28, edges = 38, number of communities = 4}
         \label{fig:jakarta_2}
     \end{subfigure}
     \hfill
     \begin{subfigure}[b]{0.4\textwidth}
         \centering
         \includegraphics[scale=0.2]{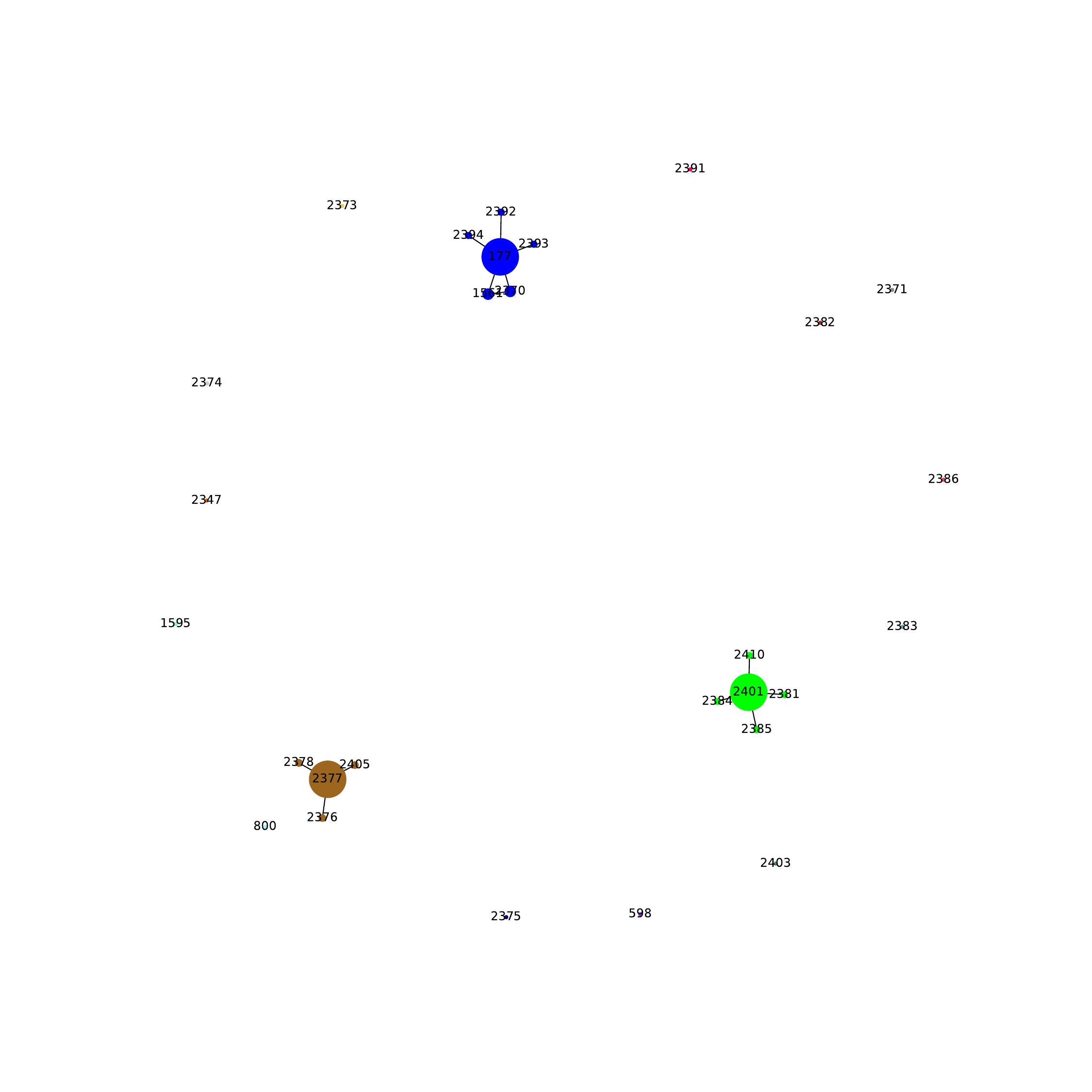}
         \caption{snapshot 3 : nodes = 28, edges = 13, number of communities = 16}
         \label{fig:jakarta_3}
     \end{subfigure}
     \hfill
     \begin{subfigure}[b]{0.4\textwidth}
         \centering
         \includegraphics[scale=0.2]{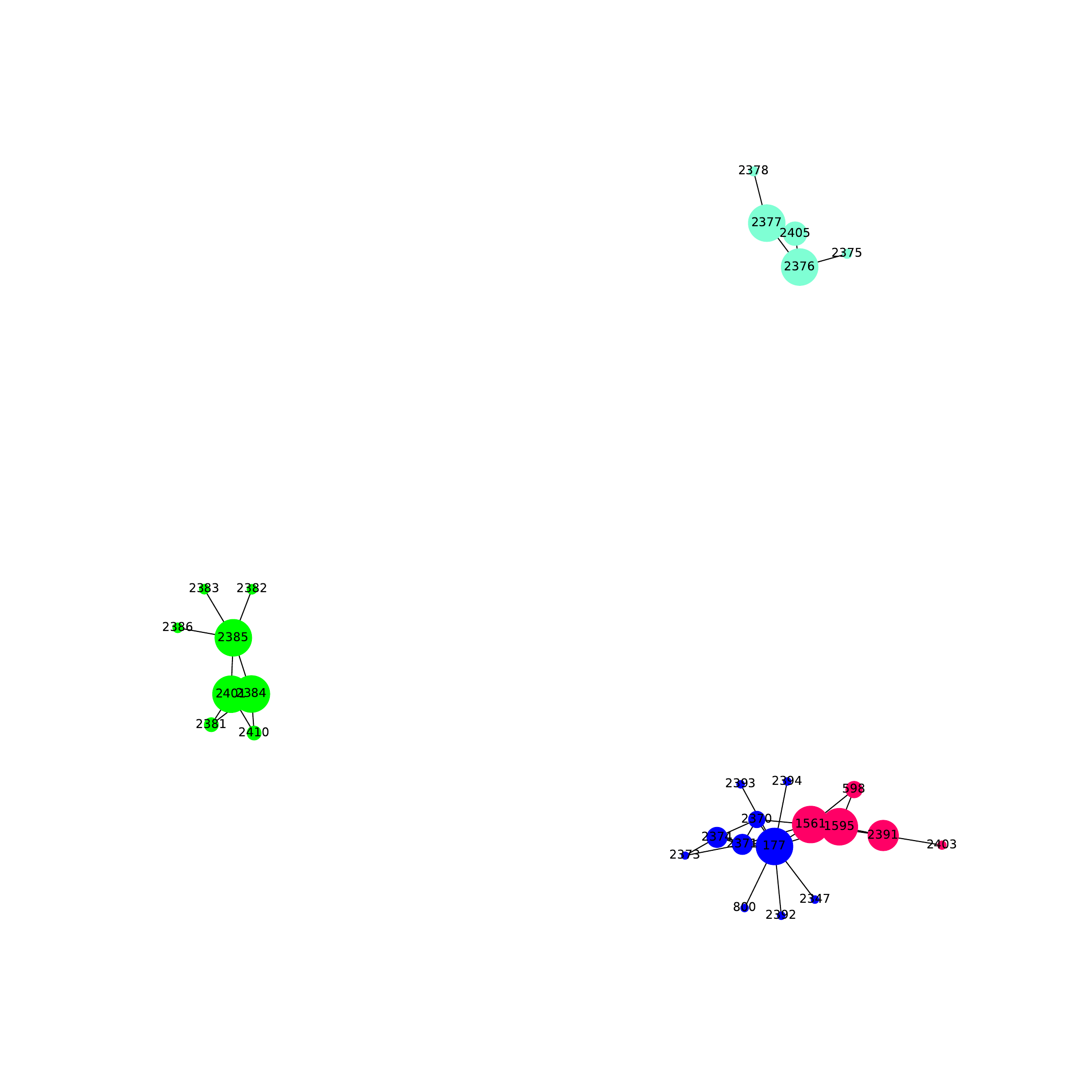}
         \caption{snapshot 4 : nodes = 28, edges = 38, number of communities = 4}
         \label{fig:jakarta_4}
     \end{subfigure}
     \hfill
        \caption{Community closeness in Jakarta terrorists network}
        \label{fig:jakarta}
\end{figure}     

 \begin{figure}[h]
     \centering
     \begin{subfigure}{0.21\textwidth}
         \centering
         \includegraphics[scale=0.09]{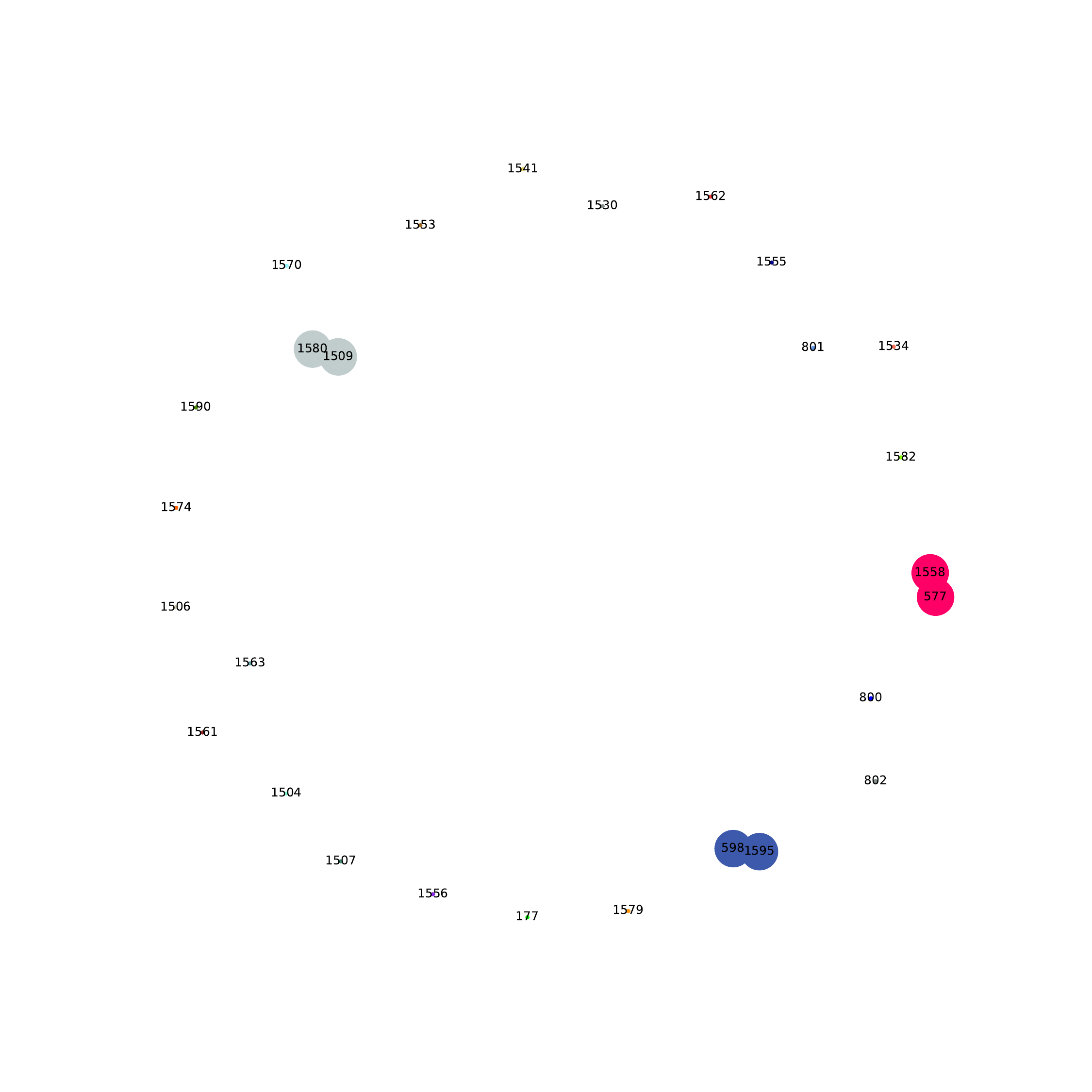}
         \caption{snapshot 1: nodes=27,edges=3,no. of communities=24}
         \label{fig:jakarta_1}
     \end{subfigure}
     \hfill
     \begin{subfigure}{0.21\textwidth}
         \centering
         \includegraphics[scale=0.09]{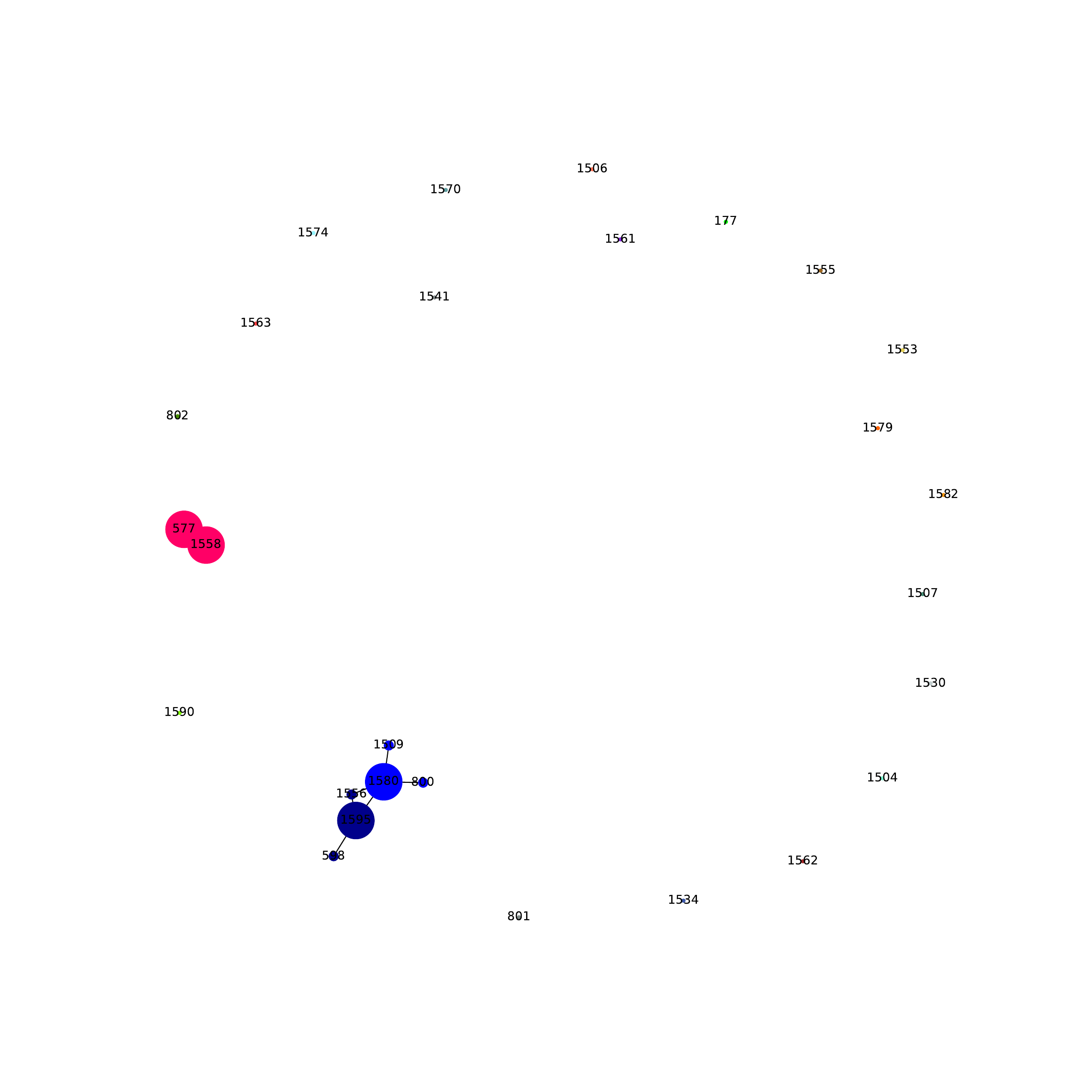}
         \caption{snapshot 2: nodes=27,edges=7,no. of communities=22}
         \label{fig:jakarta_2}
     \end{subfigure}
     \hfill
     \begin{subfigure}{0.21\textwidth}
         \centering
         \includegraphics[scale=0.09]{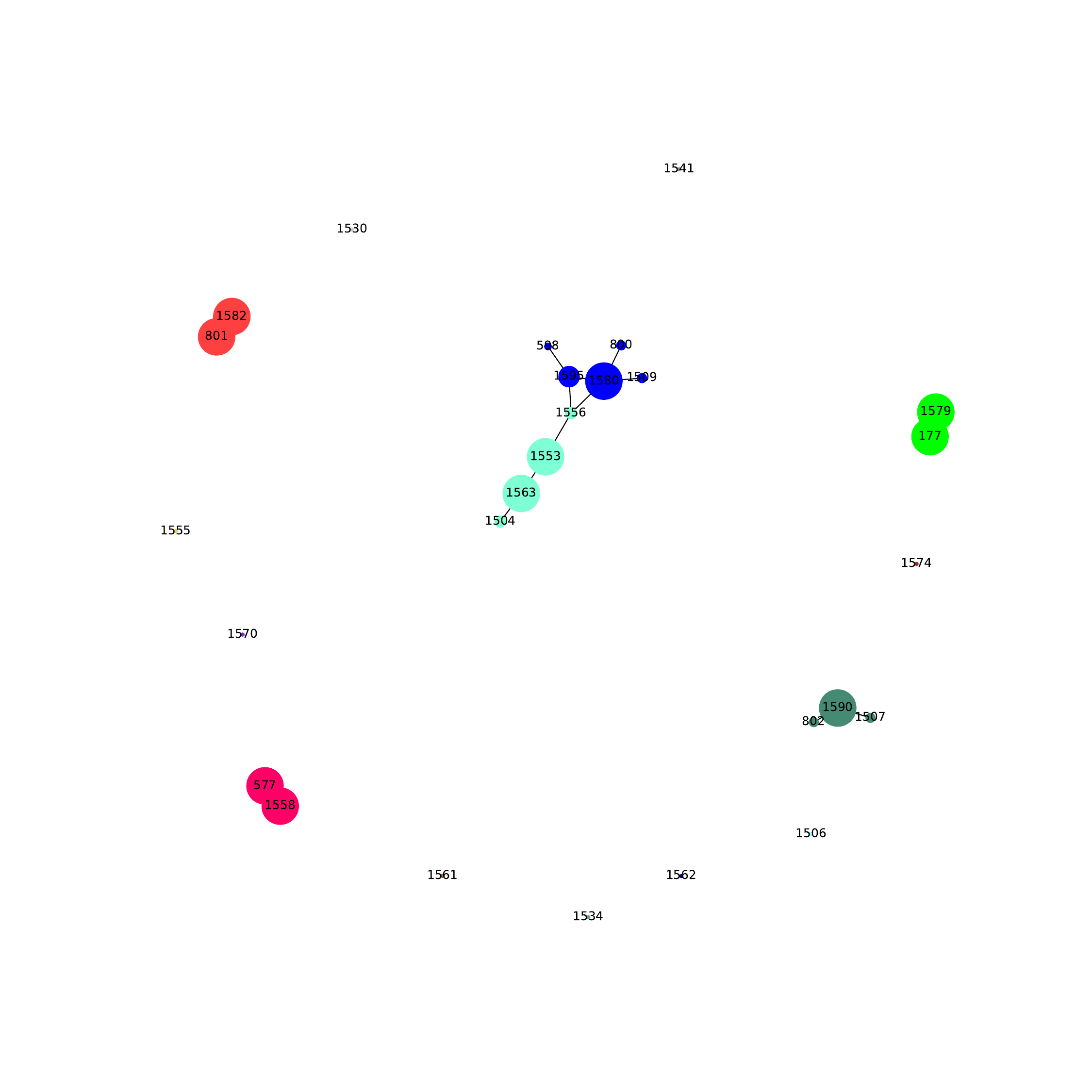}
         \caption{snapshot 3: nodes=27,edges=14,no. of communities=15}
         \label{fig:jakarta_3}
     \end{subfigure}
     \hfill
     \begin{subfigure}{0.21\textwidth}
         \centering
         \includegraphics[scale=0.09]{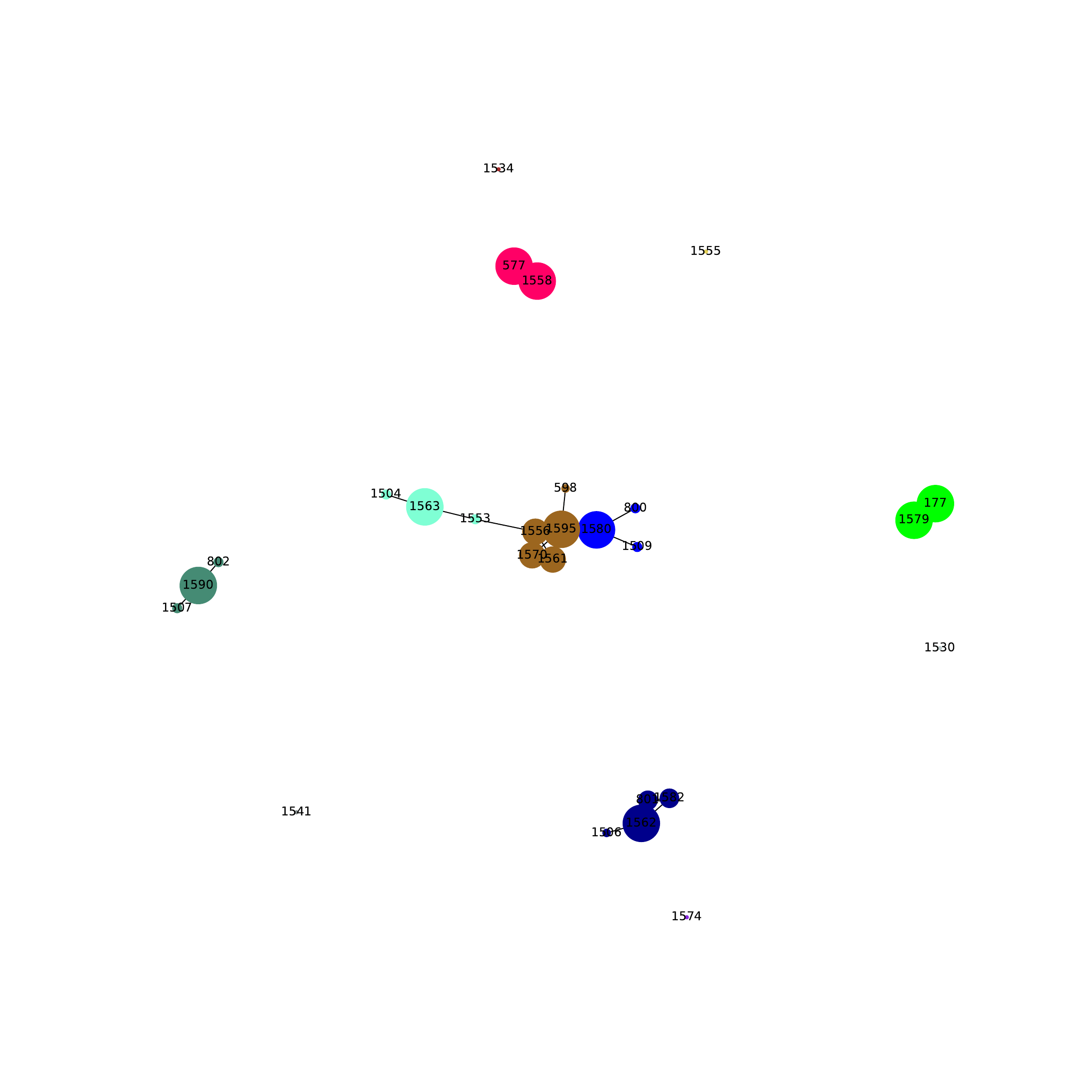}
         \caption{snapshot 4: nodes=27,edges=22,no of communities=12}
         \label{fig:jakarta_4}
     \end{subfigure}
     \hfill
      \begin{subfigure}[b]{0.21\textwidth}
         \centering
         \includegraphics[scale=0.09]{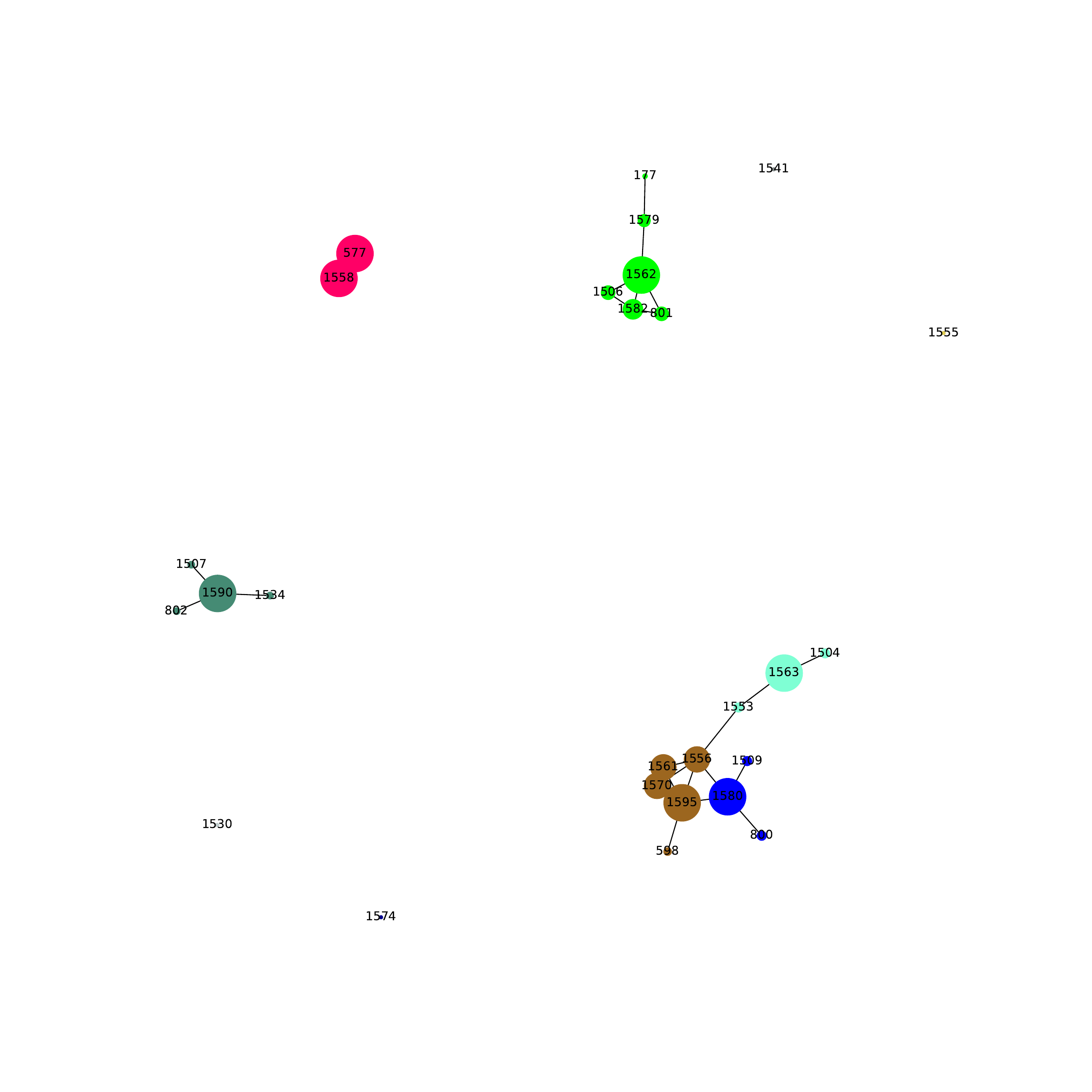}
         \caption{snapshot 5: nodes=27,edges=25,no. of communities=10}
         \label{fig:jakarta_1}
     \end{subfigure}
     \hfill
     \begin{subfigure}[b]{0.21\textwidth}
         \centering
         \includegraphics[scale=0.09]{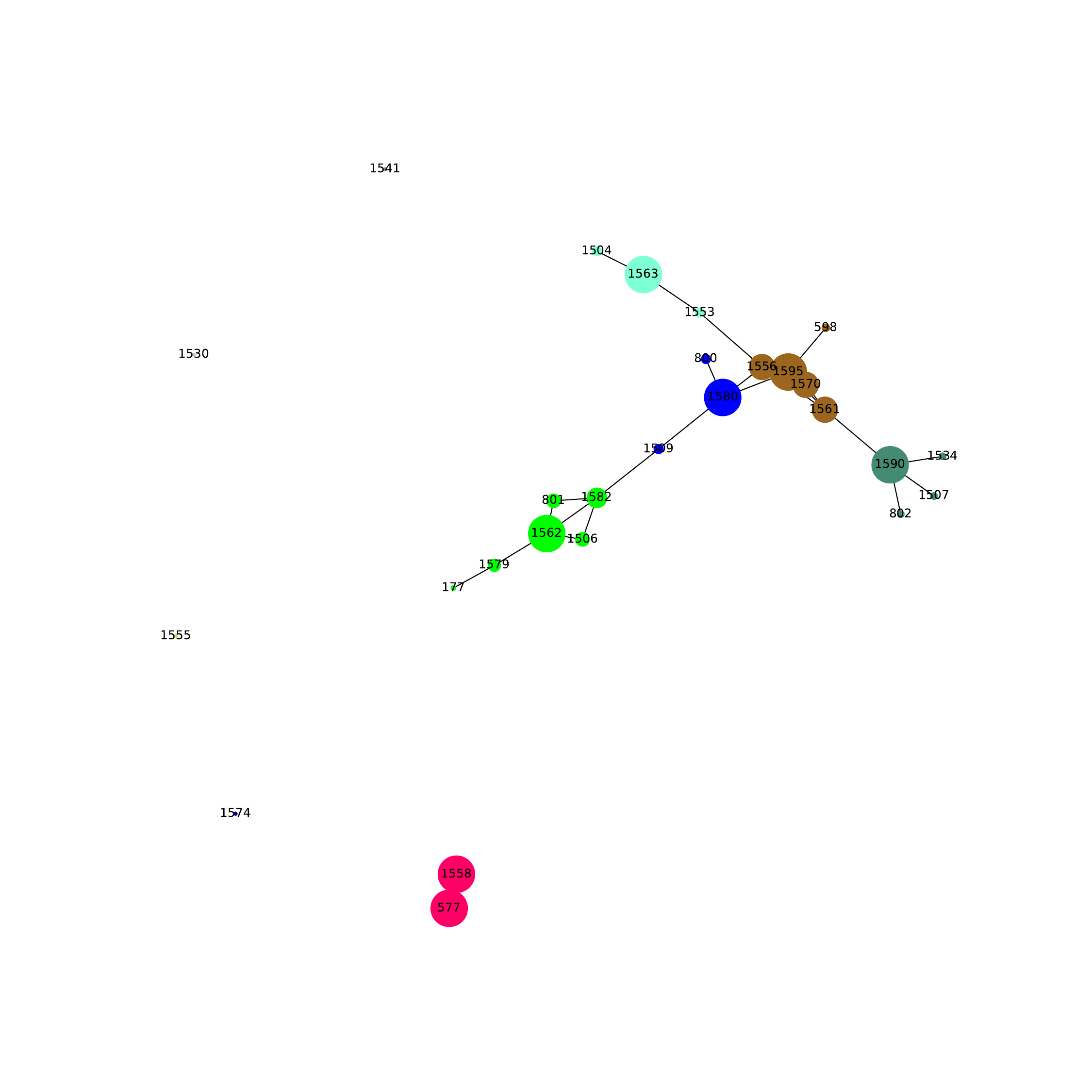}
         \caption{snapshot 6: nodes=27,edges=27,no. of communities=10}
         \label{fig:jakarta_2}
     \end{subfigure}
     \hfill
     \begin{subfigure}[b]{0.21\textwidth}
         \centering
         \includegraphics[scale=0.09]{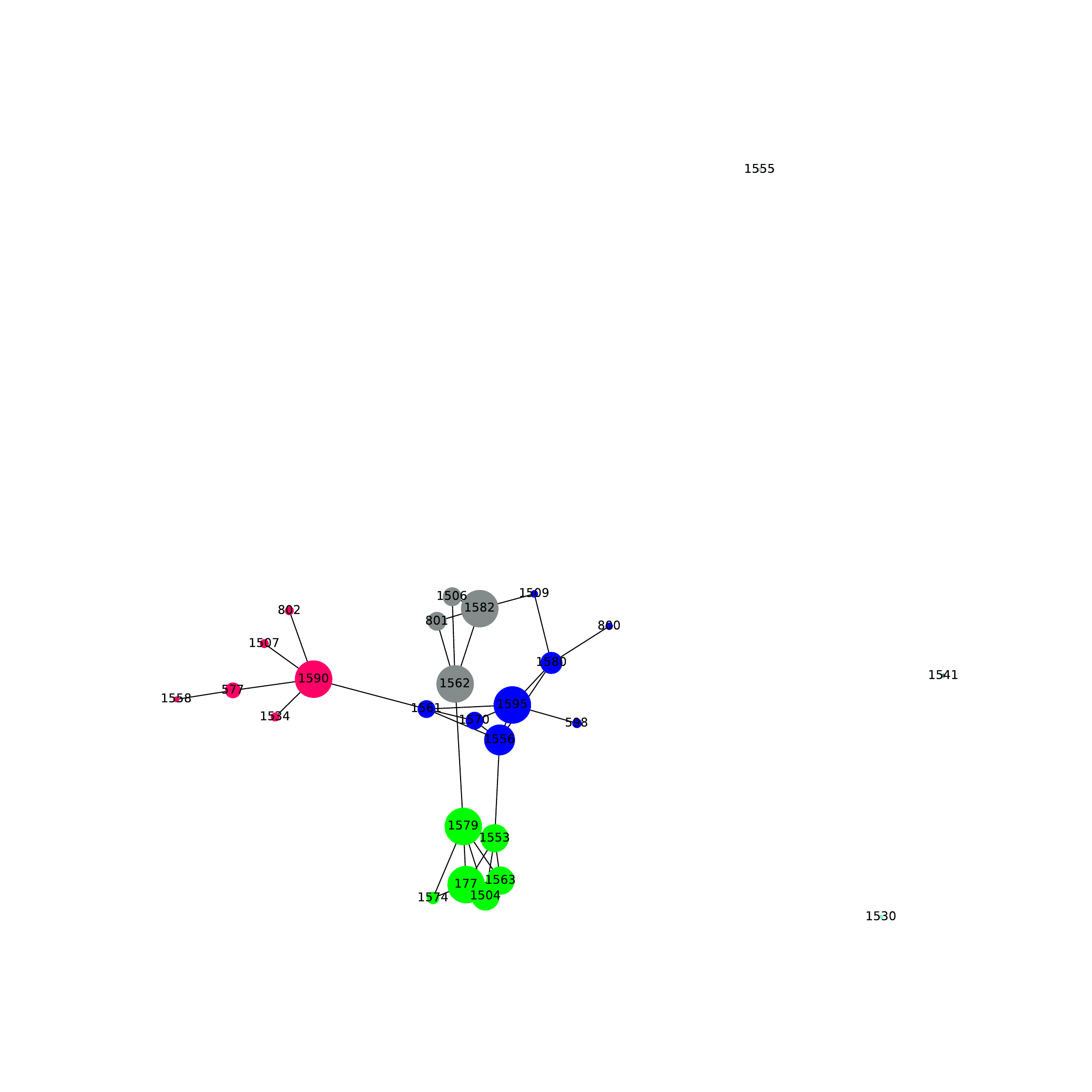}
         \caption{snapshot 7: nodes=27,edges=37,no. of communities=10}
         \label{fig:jakarta_3}
     \end{subfigure}
     \hfill
     \begin{subfigure}[b]{0.21\textwidth}
         \centering
         \includegraphics[scale=0.09]{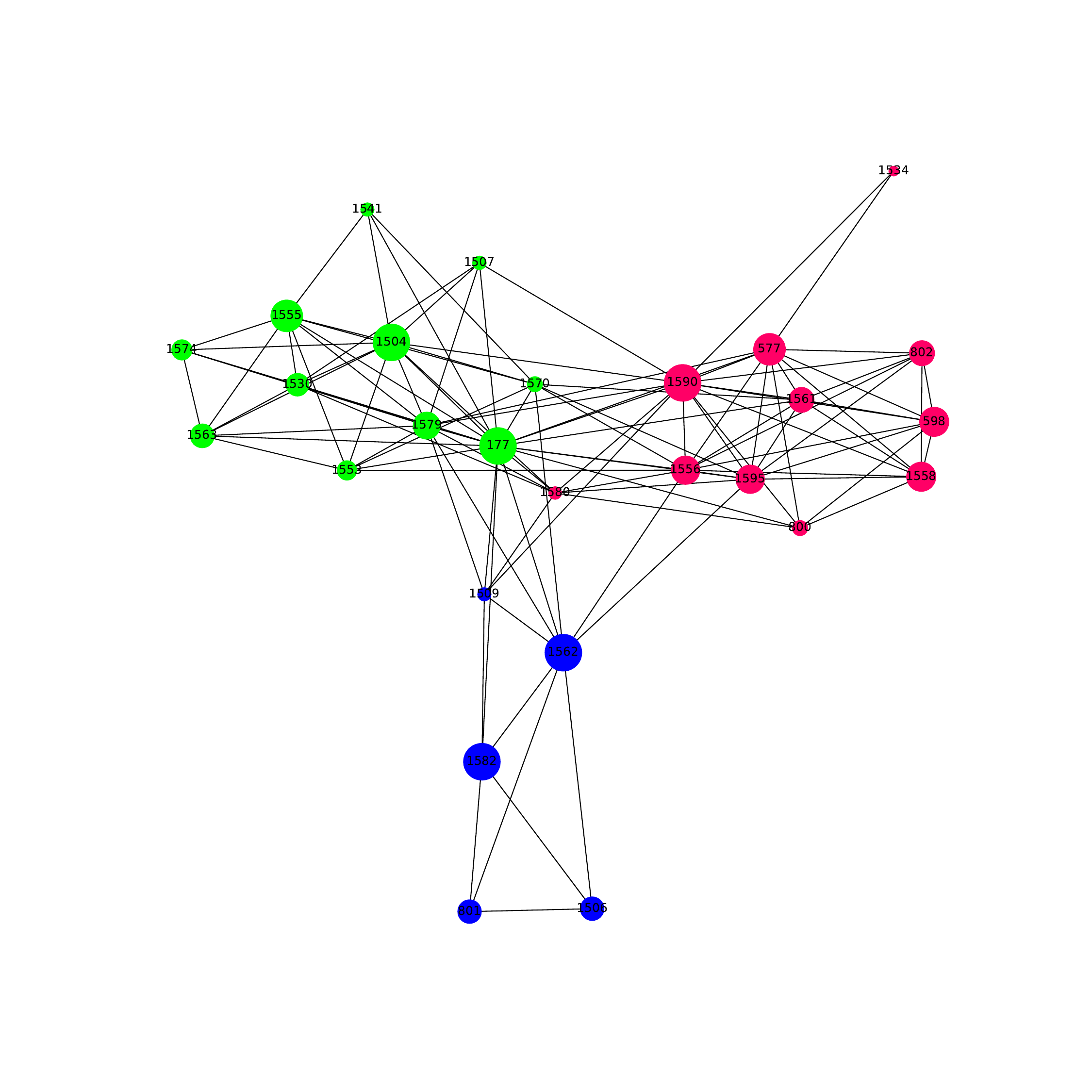}
         \caption{snapshot 8: nodes=27,edges=112,no. of communities=3}
         \label{fig:jakarta_4}
     \end{subfigure}
     \begin{subfigure}[b]{0.21\textwidth}
         \centering
         \includegraphics[scale=0.09]{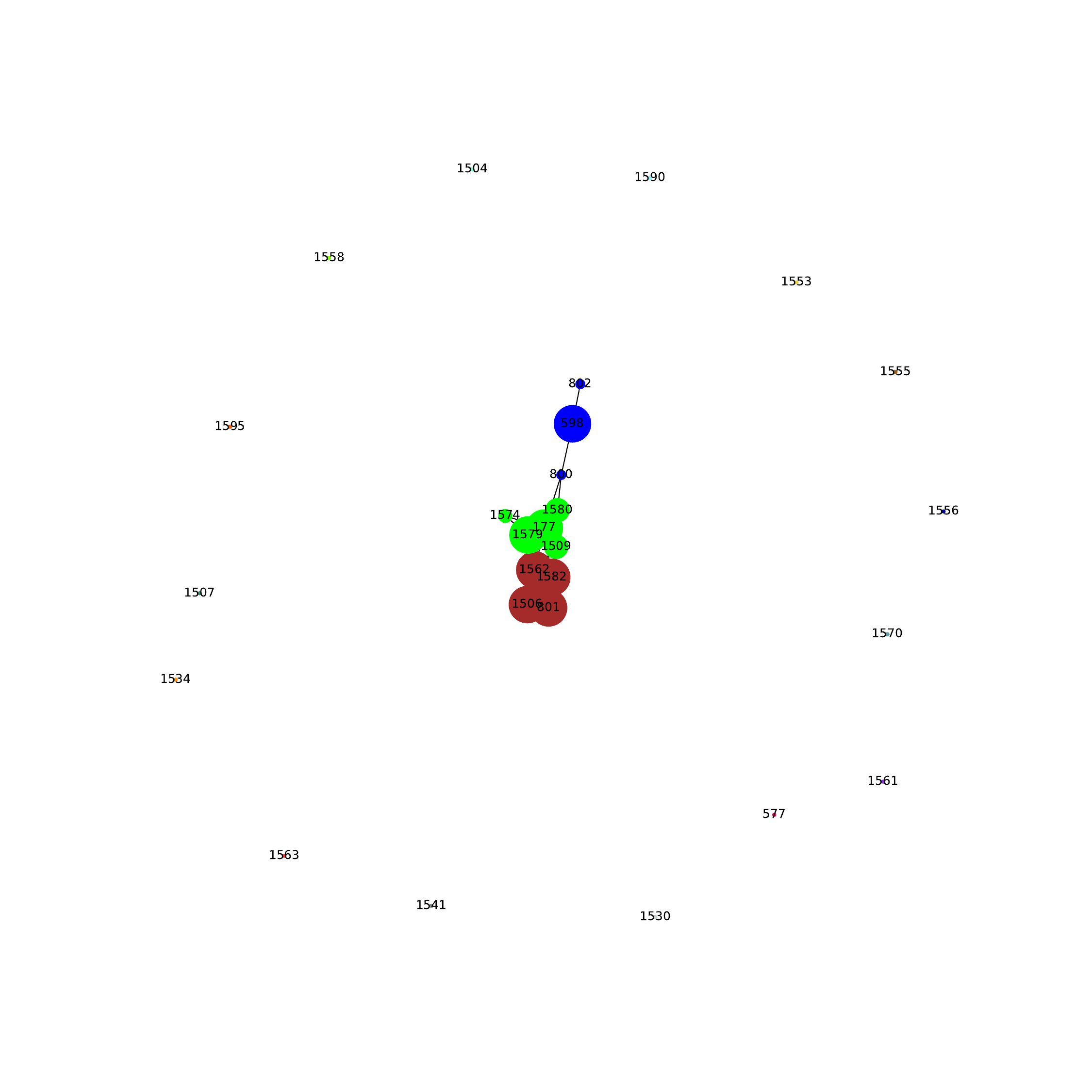}
         \caption{snapshot 9: nodes=27,edges=23,no. of communities=18}
         \label{fig:jakarta_1}
     \end{subfigure}
     \hfill
     \begin{subfigure}[b]{0.21\textwidth}
         \centering
         \includegraphics[scale=0.09]{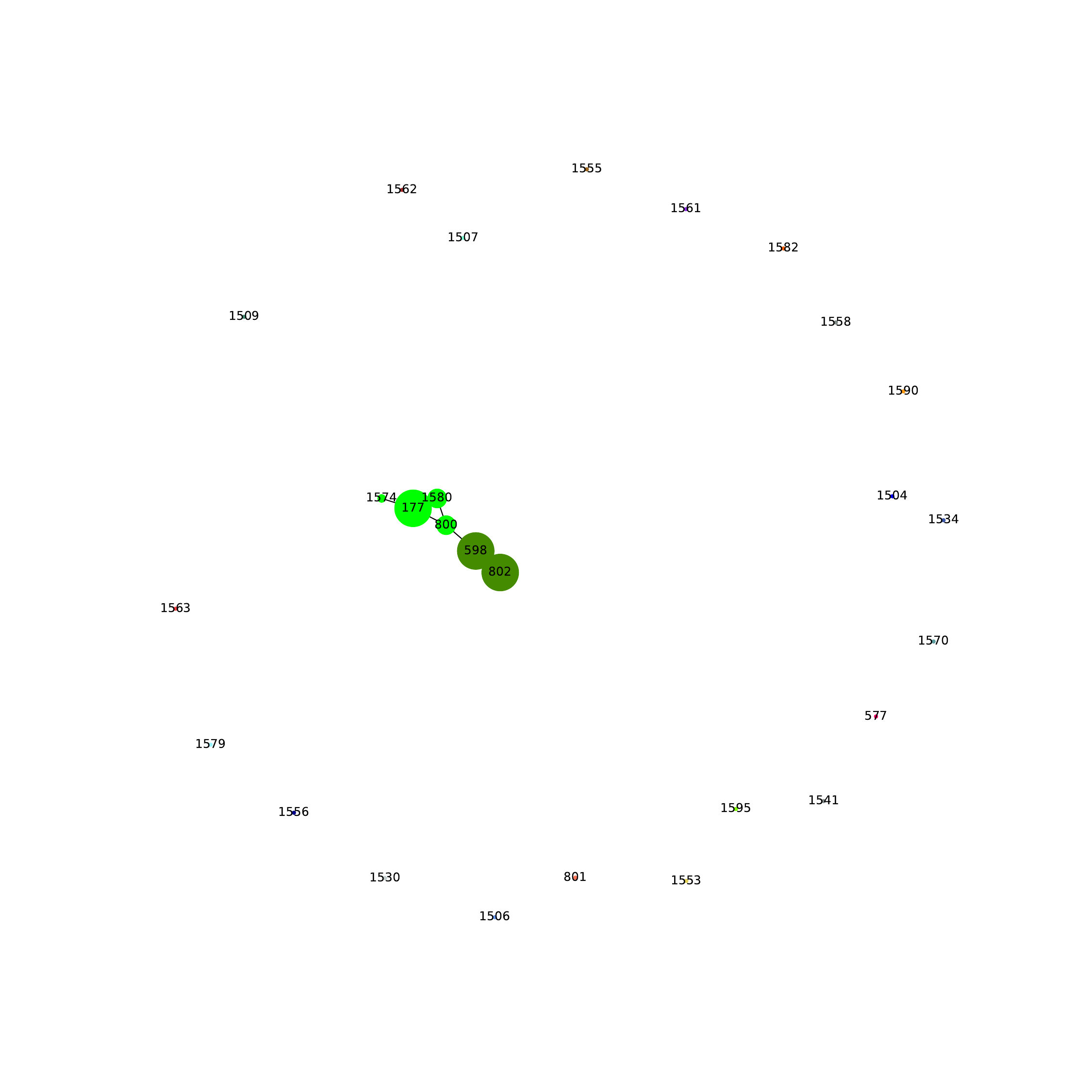}
         \caption{snapshot 10: nodes=27,edges=6,no. of communities=23}
         \label{fig:jakarta_2}
     \end{subfigure}
     \hfill
     \begin{subfigure}[b]{0.21\textwidth}
         \centering
         \includegraphics[scale=0.09]{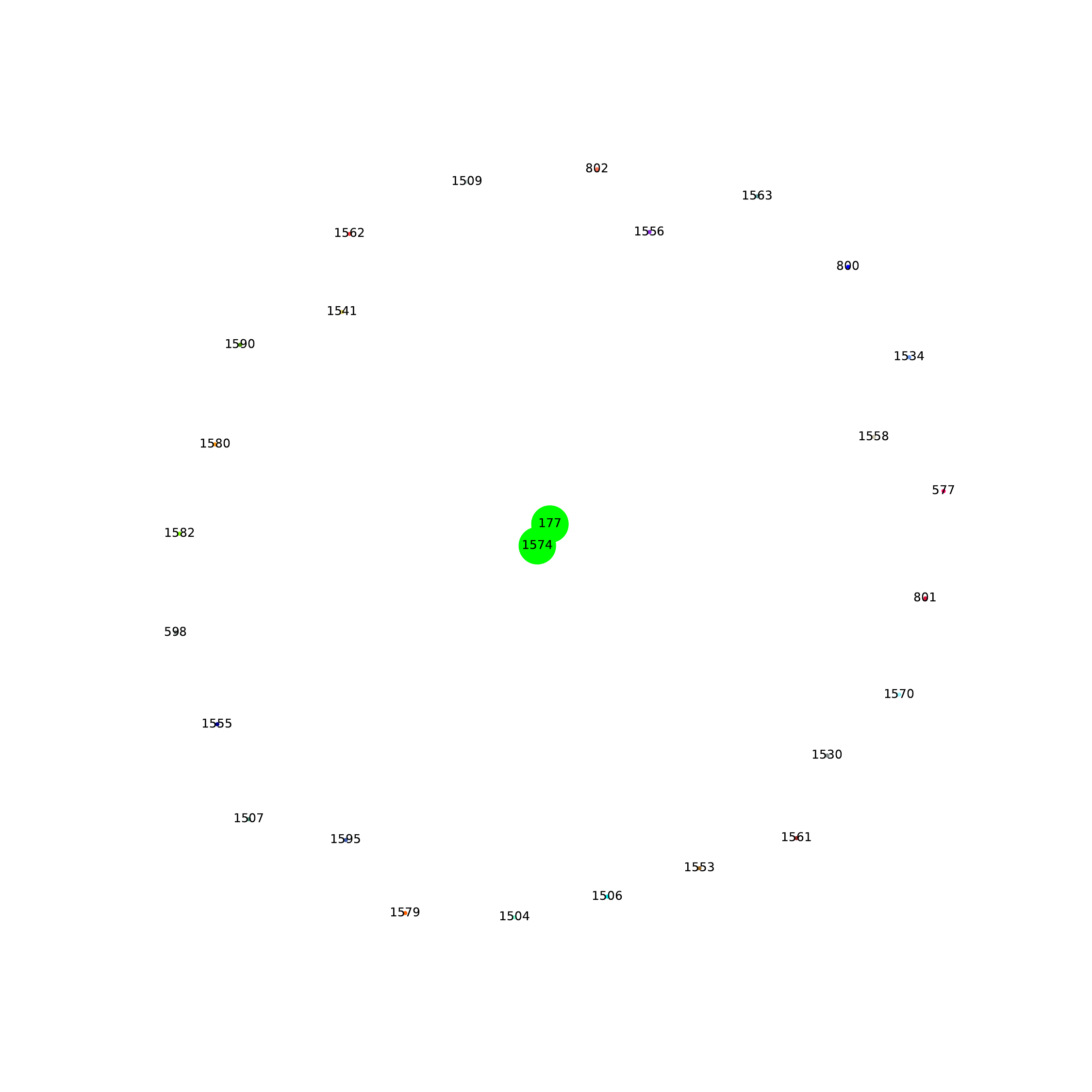}
         \caption{snapshot 11: nodes=27,edges=1,no. of communities=26}
         \label{fig:jakarta_3}
     \end{subfigure}
     \hfill
     \begin{subfigure}[b]{0.21\textwidth}
         \centering
         \includegraphics[scale=0.09]{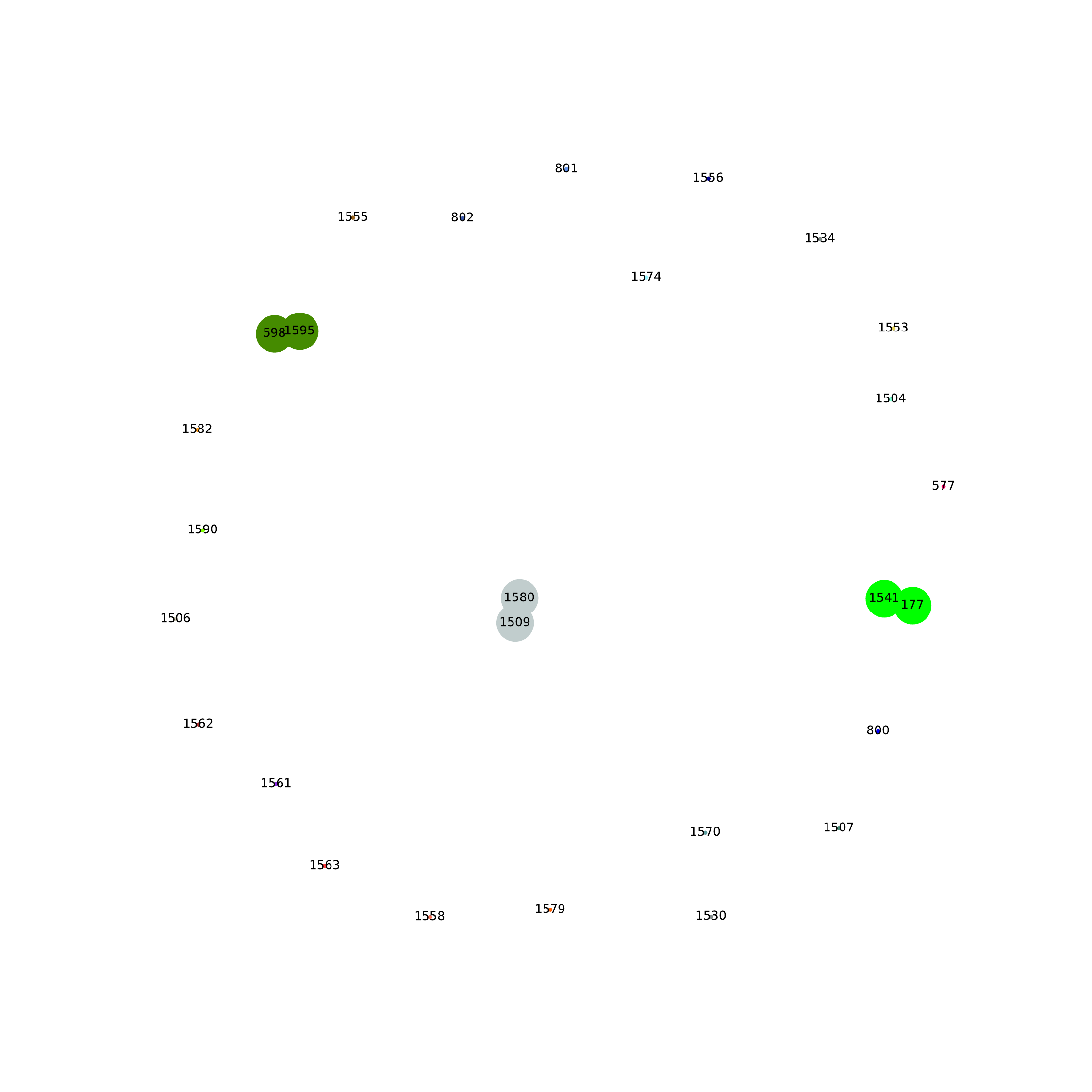}
         \caption{snapshot 12: nodes=27,edges=3,no. of communities=24}
         \label{fig:jakarta_4}
     \end{subfigure}
     \hfill
      \begin{subfigure}[b]{0.21\textwidth}
         \centering
         \includegraphics[scale=0.09]{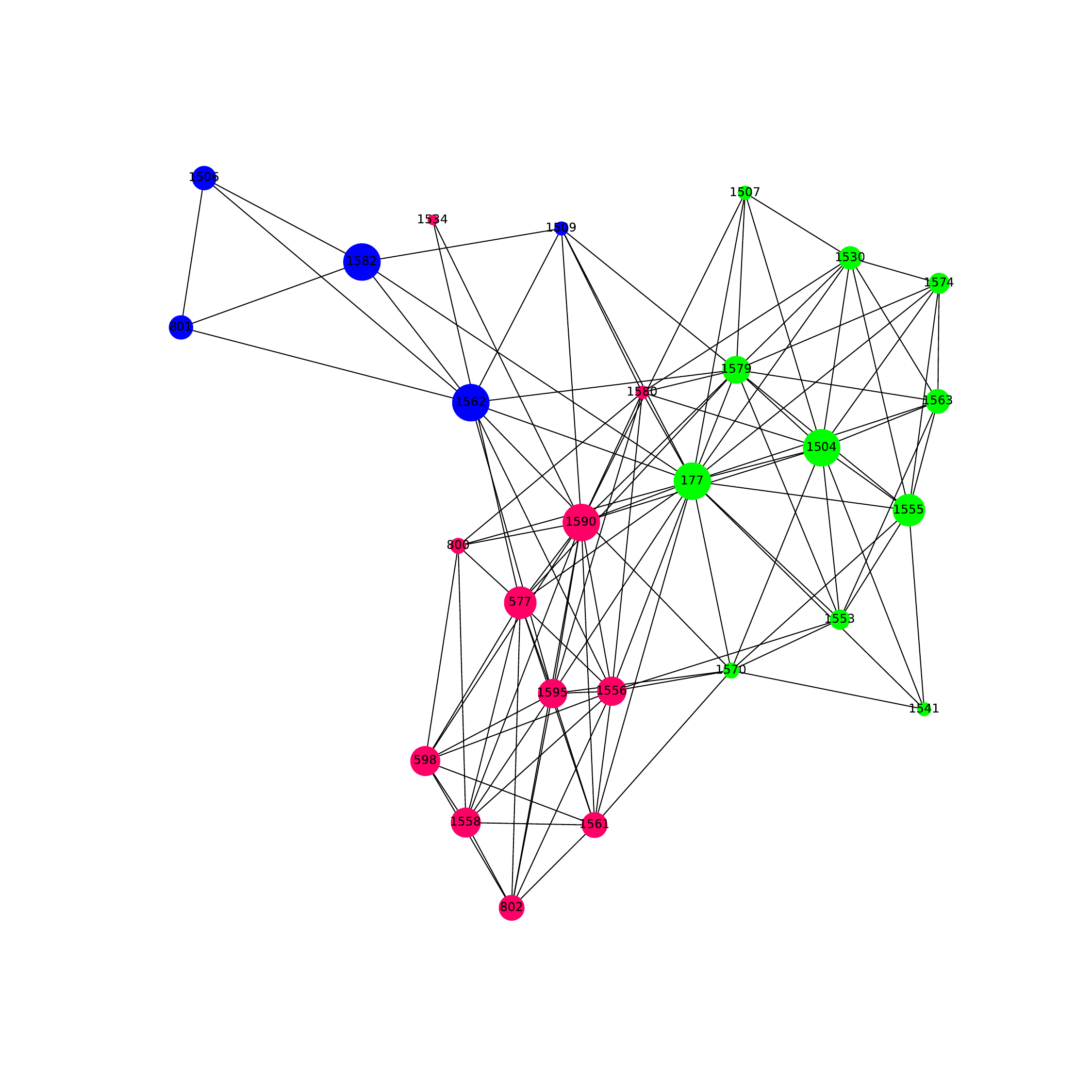}
         \caption{snapshot 13: nodes=27,edges=112,no. of communities=3}
         \label{fig:jakarta_1}
     \end{subfigure}
     \hfill
     \begin{subfigure}[b]{0.21\textwidth}
         \centering
         \includegraphics[scale=0.09]{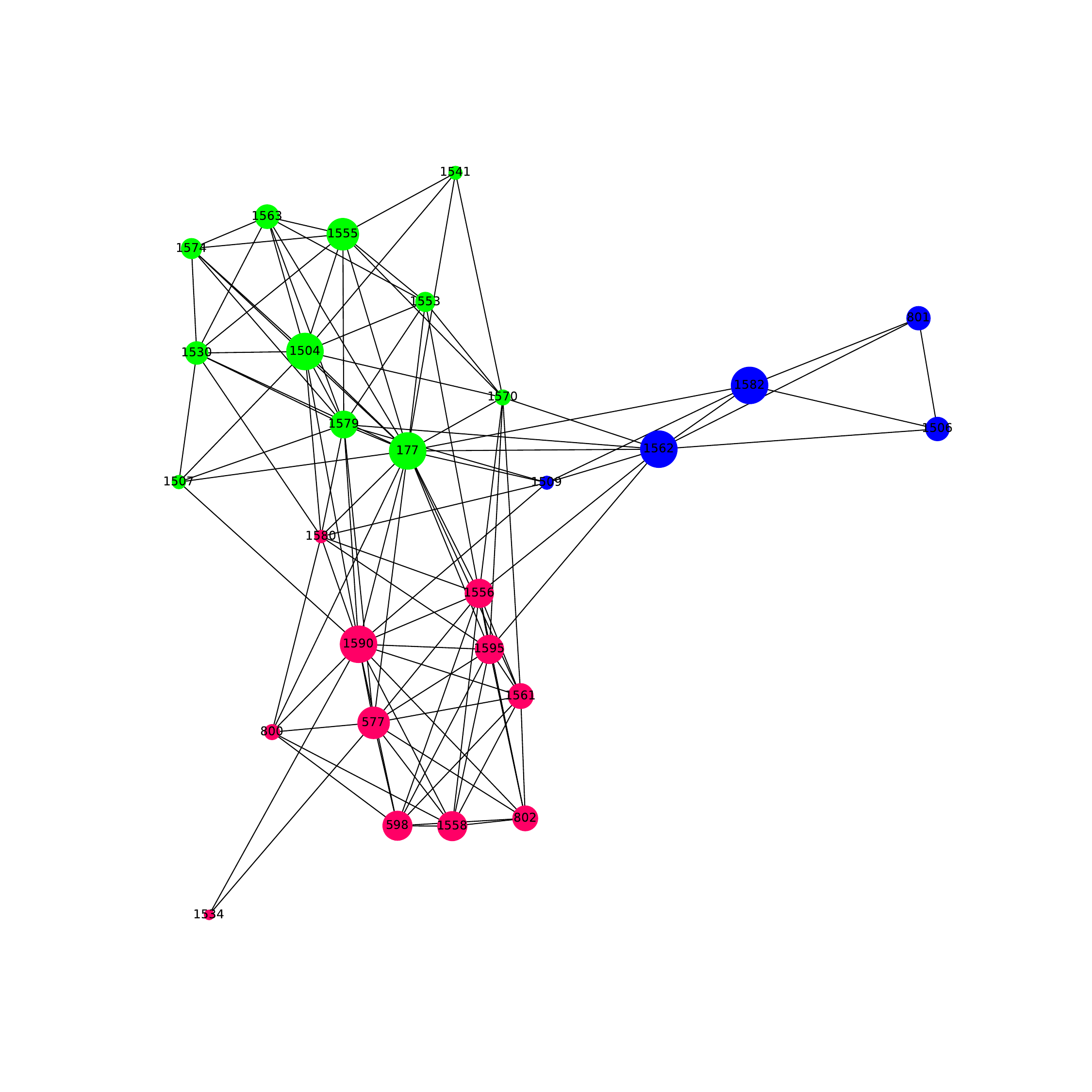}
         \caption{snapshot 14: nodes=27,edges=112,no. of communities=3}
         \label{fig:jakarta_2}
     \end{subfigure}
     \hfill
     \begin{subfigure}[b]{0.21\textwidth}
         \centering
         \includegraphics[scale=0.09]{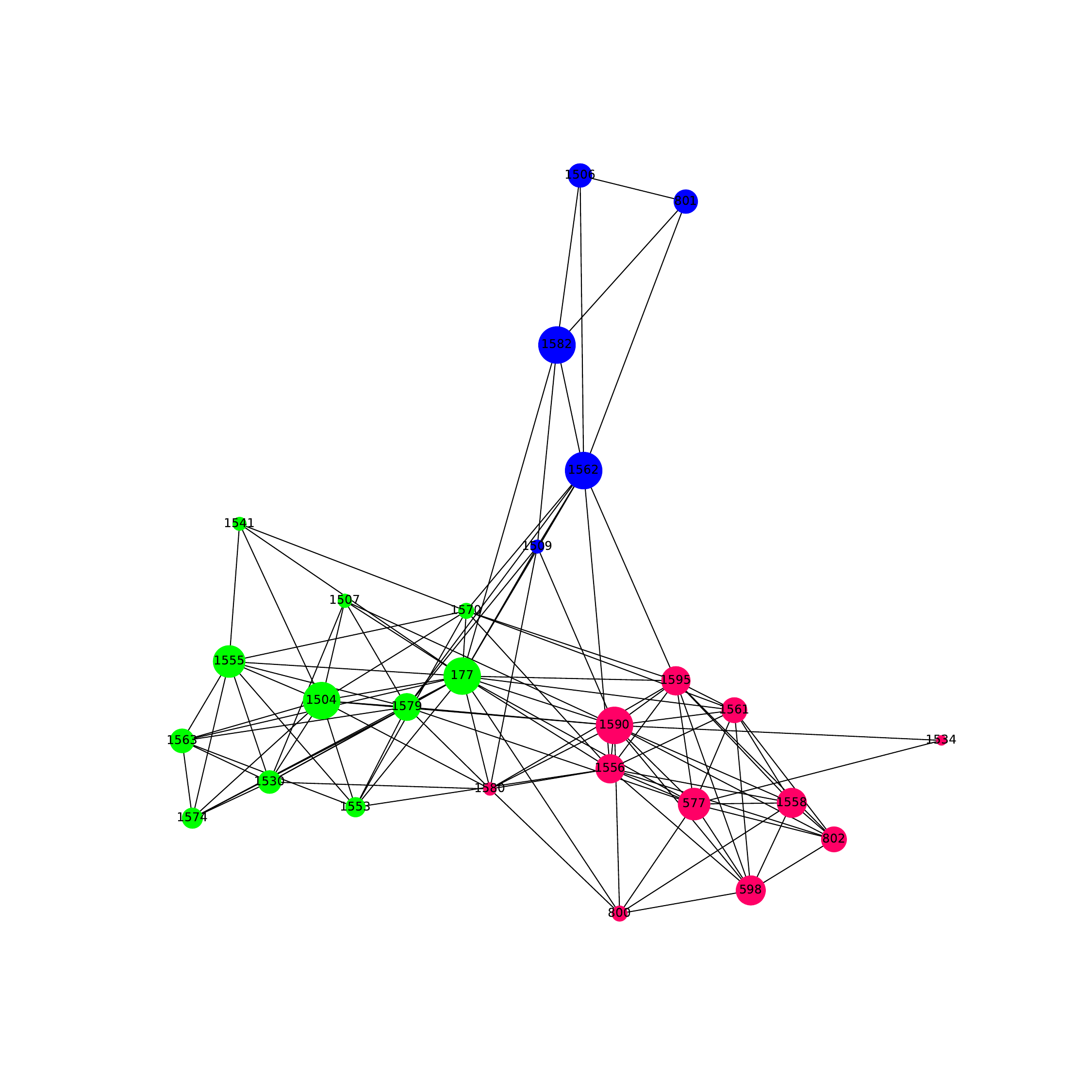}
         \caption{snapshot 15: nodes=27,edges=112,no. of communities=3}
         \label{fig:jakarta_3}
     \end{subfigure}
     \hfill
        \caption{Community closeness in Australian Embassy bombing network}
        \label{fig:australian}
\end{figure}     
\end{appendices}

\end{document}